\documentclass[usenatbib, a4paper]{mn2e}
\pdfoutput=1

\usepackage{times}
\usepackage{subfigure}
\usepackage{amsmath}
\usepackage{amsfonts}
\usepackage[pdftex]{graphicx}
\usepackage{multirow}
\usepackage{pbox}
\usepackage{hyperref}
\usepackage{url}
\usepackage{bm}
\usepackage{ulem}
\usepackage[abs]{overpic}

\renewcommand{\d}[1]{\ensuremath{\operatorname{d}\!{#1}}}
\newcommand {\apgt} {\ {\raise-.5ex\hbox{$\buildrel>\over\sim$}}\ }
\newcommand {\aplt} {\ {\raise-.5ex\hbox{$\buildrel<\over\sim$}}\ }

\title[21-cm statistics during reionization]{Distinguishing models of reionization using future radio observations of 21-cm 1-point statistics} 
\author[C. A. Watkinson \& J. R. Pritchard]
{C.~A.~Watkinson$^1$\thanks{Email: \href{mailto:c.watkinson11@imperial.ac.uk}{\protect\nolinkurl{c.watkinson11@imperial.ac.uk}}}\& J.~R.~Pritchard$^1$ \\ 
$^1$Department of Physics, Blackett Laboratory, Imperial College, London SW7 2AZ, UK\\
}
 
\date{\today}

\pubyear{2014}

\begin{document}
\maketitle

\begin{abstract}
We explore the impact of reionization topology on 21-cm statistics.
Four reionization models are presented which emulate large ionized 
bubbles around over-dense regions ({\small 21CMFAST}/ global-inside- out),
small ionized bubbles in over-dense regions (local-inside-out),
large ionized bubbles around under-dense regions (global-outside-in)
and small ionized bubbles around under-dense
regions (local-outside-in). We show that first generation
instruments might struggle to distinguish global models
using the shape of the power spectrum alone. All instruments
considered are capable of breaking this degeneracy with the
variance, which is higher in outside-in models. Global models
can also be distinguished at small scales from a boost in the
power spectrum from a positive correlation between the density
and neutral-fraction fields in outside-in models. Negative
skewness is found to be unique to inside-out models and we find
that pre-SKA instruments could detect this feature in maps smoothed
to reduce noise errors. The early, mid and late phases of reionization
imprint signatures in the brightness-temperature moments, 
we examine their model dependence and find pre-SKA instruments
capable of exploiting these timing constraints in smoothed maps. 
The dimensional skewness is introduced and is shown to have 
stronger signatures of the early and mid-phase timing if the
inside-out scenario is correct.
\end{abstract}

\section{Introduction}

We live in a time when we have a seemingly excellent 
understanding of the Universe in which we reside. 
The Planck experiment \citep{PlanckCollaboration2013b} find the Lambda Cold 
Dark Matter ($\Lambda$CDM) paradigm to be an excellent fit 
to observations of the cosmic microwave 
background (CMB), constraining the six parameters 
that describe this model to percent level accuracy. 
The CMB decoupled from matter 
approximately 380,000 years after the Big Bang and so is sensitive to 
the conditions of the early Universe; subsequent interactions 
with the rest of the Universe's energy content imprinted
signatures that provide insight into the evolution 
of the Universe. However, there exists a gap
in our understanding of this evolution that needs 
filling: the epoch of reionization (EoR), during
which the intergalactic medium (IGM) transitioned from being
completely neutral after the CMB decoupled to become 
completely ionized around a billion years after the Big
Bang. We refer the reader to \citet{Barkana2001} 
and \citet{Loeb2013} for a good introduction to reionization.

The reionization process is complex, sensitive to many
uncertainties such as the nature and evolution of the sources responsible,
(e.g \citealt{Robertson2013}), 
the degree to which the IGM clumped into dense neutral 
sinks of ionising radiation 
(e.g. \citealt{Miralda-Escude2000, Furlanetto2005}), 
and modulation (if any) from feedback effects
(e.g. \citealt{Efstathiou1992, Haiman2000, Dijkstra2004}); 
even the detailed timing of the overall process remains 
an open question.
Given this uncertainty, it is vital that we fully explore all
possible aspects of this process so that 
our interpretation of reionization related 
observations are well informed.

We have some observational constraints on the epoch of reionization. 
Thomson scattering of CMB photons by free electrons
produced during reionization wash out temperature 
fluctuations and induce large
scale polarization anisotropies. This provides 
constraints on the integral optical depth to the CMB last 
scattering surface, indicating
that reionization must have been under way by 
$z\sim 11$ \citep{PlanckCollaboration2013b, Bennett2012}; this assumes 
instantaneous reionization, 
which is unlikely to be a realistic scenario. 
Measurements from the thermal history of the IGM
are consistent with this figure \citep{Theuns2002,Bolton2010a,Lidz2010}.

Measurements of the Lyman-$\alpha$ forest in high redshift
quasars indicate that reionization was finished by 
$z\sim 6.5$ 
\citep{Becker2001, Fan2002, Fan2006} and the detection of a Lyman-$\alpha$
damping wing and small proximity zone in the most distant known
quasar at $z= 7.085$ is consistent with an IGM neutral 
fraction of more than 10\% \citep{Bolton2011, Mortlock2011}.  
The existing constraints from quasars have been bolstered 
by hints from an apparent deficit of Lyman-$\alpha$ 
emitters that the neutral fraction is above 10\% 
by $z\sim 7$ 
\citep{Ota2010, Pentericci2011, Ono2012, Schenker2012}.

To improve constraints on the epoch of reionization, work is under way 
to exploit the weak imprint of the H\thinspace\textsc{i} 21-cm spectral line on the CMB. 
This transition results from a spin flip in neutral hydrogen's lowest 
energy level, corresponding to a rest-frame wavelength $\lambda\approx21.1$cm 
and frequency $\nu\approx1420.4$ MHz \citep{EWEN1951, Prodell1952}. 
As is usual in radio observations, 
radiation intensity $I_\nu$ is described in terms of 
brightness temperature, $T_{\rm b}$, defined such that $I_\nu=B(T_{\rm b})$; 
$B(T)$ is the Planck black-body spectrum and is well approximated by the 
Rayleigh-Jeans formula at the frequencies 
relevant to reionization studies. 

The measured quantity is the differential brightness temperature
$\delta T_{\rm b}=T_{\rm b}-T_{\gamma}$ along a line of sight, 
where $T_{\gamma}$ is the temperature of the CMB. 
This differential brightness temperature evolves 
according to \citep{Field1958,Field1959a, Madau1997},
\begin{equation}
\begin{split}
\delta T_{\rm b}=&\frac{T_{\rm s}-T_{\gamma}}{1+z}(1-e^{-\tau_{\nu_0}})\\
\approx&\,27.\,\frac{T_{\rm s}-T_{\gamma}}{T_{\rm s}}\,x_{\rm \textsc{hi}}(1+\delta)\left[\frac{H(z)/(1+z)}{\d v_{\rm r}/\d r}\right]\\
&\times \left(\frac{1+z}{10}\frac{0.15}{\Omega_{\rm m}h^2}\right)^{1/2}\left(\frac{\Omega_{\rm b}h^2}{0.023}\right) \rm mK
\,.\\ \label{eqn:brightTemp}
\end{split}
\end{equation} 

\noindent This quantity is dependent on the neutral 
fraction of hydrogen $x_{\rm \textsc{hi}}$,
the matter overdensity $\delta$, line of sight velocity
gradient $\d v_{\rm r}/\d r$,
Hubble parameter $H(z)$,
matter density parameter $\Omega_{\rm m}$, 
and baryon 
density parameter $\Omega_{\rm b}$; $\Omega_i=\rho_i/\rho_{\rm c}$ where
$\rho_{\rm c}$ is the critical density required for
flat universe. 
The differential temperature also depends 
on the gas spin temperature $T_{\rm s}$, which measures 
the occupation levels of the two hyperfine 
energy levels involved in the H\thinspace\textsc{i} 21-cm line and determines 
the 21-cm optical depth $\tau_{\nu_0}$. 
We will neglect this dependence by assuming 
$T_{\rm s} \gg T_{\gamma}$, i.e. the neutral 
gas has been heated well above the CMB for the redshift range
considered \citep{Pritchard2008}. However, it is quite 
possible that spin-temperature fluctuations may be influential 
during reionization \citep{Pritchard2007}, 
we defer this question to future work.
For a comprehensive review of the 21-cm line we refer 
the reader to \citet{Furlanetto2006b, Morales2010} and \citet{Pritchard2011}.

There are many radio telescopes operating, under construction, 
or in the design stages that aim to measure the 21-cm line, e.g.
the Murchison Wide-field Array\footnote{\url{http://www.mwatelescope.org/}}
(MWA), the LOw Frequency ARray\footnote{\url{http://www.lofar.org/}}
(LOFAR), 
the Precision Array to Probe Epoch of Reionization
\footnote{\url{http://eor.berkeley.edu/}} (PAPER), 
the Giant Meter-wave Radio Telescope
\footnote{\url{http://gmrt.ncra.tifr.res.in/}} (GMRT), 
and the Square Kilometre Array
\footnote{\url{http://www.skatelescope.org/}} (SKA). 
All aim to 
detect radio fluctuations in the redshifted 21-cm signal that
correspond to varying levels of neutral hydrogen. The signal 
from reionization will only be probed at a statistical level 
by pre-SKA telescopes.
Next generation instruments, such as SKA,
will make detailed maps of the 21-cm signal, enabling the measurement
of hydrogen's properties out to $z=25$. Such observations will constrain
the properties of the intergalactic medium (IGM) and therefore the 
cumulative effect of all  sources of light.

There has been much focus on the statistics of the 21-cm signal 
produced by the process of reionization, along with their 
detectability by future observations; previous studies 
have looked at the power spectrum (e.g. 
\citealt{Furlanetto2004a, Zaldarriaga2004, 
Mellema2006, Lidz2008, Pritchard2008, Signal2010, Friedrich2011}), 
the probability density function (PDF) and its moments (e.g. 
\citealt{Furlanetto2004, Wyithe2007a, Harker2009, Ichikawa2010}), 
and the difference 
PDF (e.g. \citealt{Barkana2008, Pan2012}).
This is mainly because we will be limited to statistical 
measures in the near future, but even with the precision 
of SKA, statistics will play a vital role in 
connecting theory to observation.

Despite the consensus of different observations that 
we are observing the end of reionization at $z>6$, the 
effect of absorption systems casts uncertainty on this by adjusting 
the mean free path at high redshifts. The nature of
such absorbers could also
strongly affect the morphology and environment of the ionized 
regions throughout reionization \citep{Crociani2010,Alvarez2012}.

In this work we 
test the sensitivity of various statistics
to different morphological characteristics of ionized regions
that might have formed during reionization.
We take {\small 21CMFAST} \citep{Signal2010} as 
our reference model, the properties of
which are large ionized regions around the over-dense 
regions containing sources. We then construct three contrasting toy models: 
one with small ionized bubbles in over-dense regions; 
one with small ionized bubbles in under-dense regions; and another 
with large ionized bubbles in under-dense regions.
The truth will likely be a combination of such properties,
with large-scale inside-out behaviour contrasted by
recombination effects producing
outside-in behaviour on smaller scales, e.g \citet{Furlanetto2005}.
By examining the statistics of the extreme models
described in this work, 
we are able to tease apart the effects of each property,
and therefore understand the strengths and limitations
of each statistic.

We examine the power spectrum, 1-dimensional 
PDF, skewness and variance 
of 21-cm maps, 
as well as the maps themselves; we place 
emphasis on the model dependent differences in 
these statistics. 
To approximate instrumental effects we smooth
and re-sample the maps according to the likely instrumental resolution 
of MWA, LOFAR and SKA. Errors induced by instrumental 
noise are then approximated by assuming independent but identically 
distributed random Gaussian noise with zero mean on each pixel and 
propagating these brightness-temperature errors on to our statistics.
Finally, the variability in the statistics caused by cosmic variance 
is estimated by sub-sampling.

This rest of this paper is structured as follows: 
in Section \ref{sec:modeldesc} we detail the four 
models we use to simulate the different reionization scenarios; 
in Section 
\ref{sec:toymodel} we construct an illustrative toy model 
for the PDF of 
the neutral-fraction maps and compare the 
moments of this toy model
with those measured from our four model simulations; 
we then move on to maps of brightness temperature, 
first looking at the statistics of clean maps in 
Section \ref{sec:CleanStats}; then, in Section 
\ref{sec:NoisyStats}, we study the
statistics of noisy maps that 
mimic instrumental effects; we look at the
limitations imposed by cosmic variance in section \ref{sec:CosVar}; 
and finally, in Section \ref{sec:Conc}
we discuss our conclusions. 
Unless stated otherwise all lengths are co-moving. 
Throughout we assume a standard $\Lambda$CDM cosmology 
as constrained by Planck, adopting $\sigma_8=0.829$,
$h=0.673$, $\Omega_{\rm m}=0.315$, $\boldsymbol{\Omega_{\Lambda}=0.685}$, 
$\Omega_{\rm b}=0.049$ and $n_{\rm s}=0.96 $ \citep{PlanckCollaboration2013b}. 

\section{Models for reionization}\label{sec:modeldesc}

We utilise four simulations to mimic the behaviour of different plausible  
ways in which ionized bubbles might have evolved with time.
Our aim here is not to produce a 
detailed and accurate simulation of reionization,
instead the qualitatively different models that we describe in this section 
allow clean separation of the effect that different 
morphological properties have on the statistics of the 
21-cm brightness temperature.
In reality, the characteristics we discuss in this chapter will not
be as extreme and will likely be present in combination 
with others.
Table \ref{tbl:modelkey} presents a summary of the physical motivation
behind each model, defined by the interplay between ionization 
and recombination.

The idea of global reionization in which over-dense
regions are ionized first is preferred 
by current theory and simulations; 
as such we use the well established semi-numerical code 
{\small 21CMFAST} \citep{Mesinger2007, Signal2010} 
to provide a density field and our 
fiducial global-inside-out model. 
We then use this density field as a skeleton upon 
which to construct global-outside-in, 
local-inside-out and local-outside-in models. 
We neglect spin-temperature fluctuations so that the 
contributors to fluctuations in the 21-cm brightness 
temperature are the neutral-fraction field $x_{\rm \textsc{hi}}$,
the velocity field $\mathrm{d} v_r/\mathrm{d} r$ 
and the density field $\delta$.
The four models we use differ only in the prescription 
used to construct the neutral-fraction field and thus have identical
density and velocity fields. 
The inside-out models have density fields that 
are anti-correlated with the neutral-fraction field, with 
the most dense regions being the least likely 
to remain neutral. In contrast, outside-in models  
have density fields correlated with the neutral
field so that regions of extreme overdensity will instead
host the most persistent neutral regions. Where there is sensitivity
to the global properties of the density field, larger 
bubbles of ionized gas will develop; in contrast if there is only localised
sensitivity then these bubbles will be much smaller.

\begin{table}
    \caption{Summary of the four models considered in this paper.}
    \begin{center}
      \begin{tabular}[p]{ll}
	\hline
         Figure key  &  Model motivation \\
        \hline
        \parbox{2.5cm}{\textbf{global-inside-out}\\
			Figure \ref{fig:barrier} :\\
			Region is ionized at first up-crossing of red barrier.}  
        & \parbox{5cm}{\textit{{\small 21CMFAST}} \\
          UV-fuelled ionized bubbles form and grow around
          the over-dense regions containing ionizing
          sources. The growth of these ionized bubbles
          drives the decrease in the average neutral fraction
          during the course of reionization}\\ 
	\hline
        \parbox{2.5cm}{\textbf{local-inside-out}\\
			Figure \ref{fig:barrier} :\\
			Pixel is ionized if its over-density is above blue threshold.}
        & \parbox{5cm}{\textit{pixel-by-pixel inversion 
            	of MHR00}\\
          	A high density of absorbers 
          	keep the mean free path of the UV photons
	 	small; ionized regions are limited to the 
	 	immediate neighbourhood of the originator 
		galaxy of ionizing radiation.
         	The average neutral fraction decreases with 
                time as the ionizing 
                photons from more and more UV sources are able to 
                overpower the recombinations/absorbers in their individual 
                haloes.
        }\\
        \hline
        \parbox{2.5cm}{\textbf{global-outside-in}\\
			Figure \ref{fig:barrier} :\\
			Region is ionized at first down-crossing of dotted black barrier.}
        & \parbox{5cm}{\textit{Inversion of {\small 21CMFAST}} \\
          High recombination rates in over-dense regions 
          mean that stable ionized bubbles form in the under-dense
          regions. An increasingly intense background of hard radiation
          overwhelms
          recombination rates on the higher density outskirts of 
          these bubbles and they grow, decreasing the average
          neutral fraction with time.} \\
        \hline
        \parbox{2.5cm}{\textbf{local-outside-in}\\
			Figure \ref{fig:barrier} :\\
		 	Pixel is ionized if  its over-density is below green threshold.}
        & \parbox{5cm}{\textit{pixel-by-pixel implementation 
            of MHR00} \\
          Recombinations and/or absorbers dominate and 
          only under-dense regions in very localised regions become
          ionized. The average 
          neutral fraction decreases with time as increasingly
          intense background of hard radiation
          overpowers the recombinations/absorbers in more and 
          more regions.}\\        
        \hline        
      \end{tabular}
    \end{center}
    \label{tbl:modelkey}
\end{table}

\subsection{Global-inside-out: {\small 21CMFAST}} \label{sec:21cmfast}

{\small 21CMFAST} numerically implements the approach of \citet{Furlanetto2004a}, 
henceforth FZH04, this paper applied the excursion set formalism to
the problem of reionization. 

\begin{figure}
  \centering
  \includegraphics{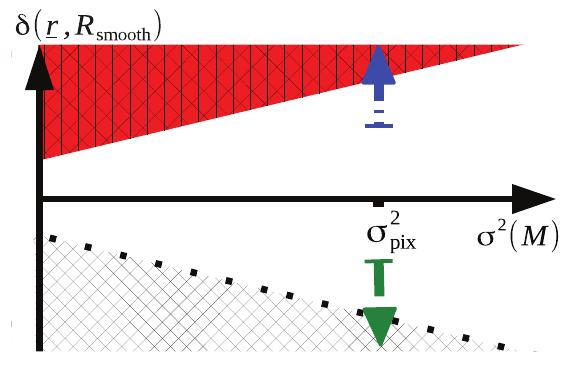}\\
  \caption{Illustration of the barriers for ionization
in overdensity against variance on a given smoothing scale
$\sigma^2(R_{\rm smooth})$. Barriers relate to
global-inside-out (solid red line with hatching above), 
local-inside-out (blue dotted arrow), 
global-outside-in (black dashed line with hatching below) and
local-outside-in (green dashed arrow).} 
  \label{fig:barrier}
\end{figure}

The excursion set model considers the evolution of the  
matter overdensity field 
$\delta(\bmath{r},R_{\rm smooth})$, a 
function of position vector $\bmath{r}$ 
and smoothing scale $R_{\rm smooth}$, with its variance
$\sigma^2(R_{\rm smooth})$.
FZH04 
constructed a barrier for ionization in overdensity as
a function of smoothing scale, illustrated 
by the red line in Figure \ref{fig:barrier}; the hatched area
above depicts the parameter space
for which a region would be ionized. 
The first up-crossing of this barrier 
(when smoothing from large to small scales) is related to 
an ionized region of corresponding mass and scale.
To construct this barrier they make the 
simple ansatz that a region will be fully ionized 
 if enough photons have been produced to ionize 
every baryon in that region, i.e. it is able to self 
ionize. The mass of an ionized 
region $m_{\rm ion} $ is then related to the mass in collapsed
 objects $m_{\rm gal}$ by an 
efficiency factor $\zeta$
according to,

\begin{equation}
m_{\rm ion} = \zeta m_{\rm gal}\,.\label{eq:FZHanz}
\end{equation}

\noindent The efficiency factor parametrizes the 
efficiency at which ionizing photons will escape a collapsed 
object, which depends on uncertain source properties 
and could be a function of time. An insightful interpretation 
is to write $\zeta=f_{\rm esc}f_*N_{\gamma/b}(1+n_{\rm rec})^{-1}$,
where $f_{\rm esc}$ is the fraction of ionizing photons that 
escape the object, $f_*$ is the star formation efficiency of the 
object, $N_{\gamma/b}$ denotes the number of ionizing photons
produced per baryon in stars, and finally $n_{\rm rec}$ is the 
typical number of times that hydrogen will have recombined.
FZH04 used the 
assumption of equation \ref{eq:FZHanz} to set up 
a condition for ionization by requiring that
a fully ionized 
region have a collapsed fraction $f_{\rm coll}\ge\zeta^{-1}$;
the collapsed fraction can be thought of as the fraction of
a region that contains ionizing sources.
Rearranging 
the extended Press--Schechter prediction for 
$f_{\rm coll}$ \citep{Bond1991,LaceyCedric1993} 
gives the ionization barrier

\begin{equation}
\begin{split}
\delta_m\ge\delta_x(m,z)\equiv\delta_c(z)-\sqrt{2}K(\zeta)
\left[\sigma_{\rm min}^2-\sigma^2_m\right]^{1/2}\,,
\end{split}
\label{eqn:FZHbarrier}
\end{equation}

\noindent in which  
$K(\zeta)=\operatorname{erf}^{-1}(1-\zeta^{-1})$,
$\delta_{\rm c}(z)$ is the critical density for gravitational collapse, 
$\sigma^2_m$ is the variance of density fluctuation 
on the scale $m(R_{\rm smooth})$, 
$\delta_{m}$ is the mean overdensity in the region,
 and $\sigma^2_{\rm min}=\sigma^2(m_{\rm min})$,
where $m_{\rm min}$ is the minimum mass for collapse.

The key implication of this model is that for every 
redshift there is a preferred scale for ionized regions.
This occurs because there is a `sweet spot' for first crossings;
a positive gradient makes the barrier easier to cross 
at large scales, where sufficiently 
large overdensities are rare, than it is to 
cross at the smaller scales where 
overdensities are more extreme. 
As the barrier moves down with 
decreasing redshift, ionized regions 
become more numerous and the characteristic 
size shifts to larger scales.

This approach was utilised
in the semi-numerical {\small 21CMFAST} 
simulation package \citep{Mesinger2007, Signal2010}. 
This generates density 
and velocity fields from Gaussian initial conditions and then approximates 
gravitational collapse using first order perturbation theory 
\citep{Zel'dovichYa.B.1970} to produce non-linear density maps.
{\small 21CMFAST} then numerically applies the excursion-set 
approach of FZH04 to construct its neutral-fraction fields 
from the density fields.
The code smooths the density field concentrically 
from a maximum allowed radius for a cohesive ionized 
region $R_{\rm max}\approx 30$Mpc
\footnote{$R_{\rm max}$ is a free parameter based on ionized
photon mean free path at the redshifts of interest, see 
\citet{Storrie-Lombardi1994, MiraldaEscude2003, Choudhury2008}}, down to that 
of a pixel $R_{\rm pix}$. At each smoothing scale it 
checks which pixels meet the condition 
$f_{\rm coll}(\bmath{r},z,R_{\rm smooth})\ge\zeta^{-1}$ 
where $\bmath{r}$ denotes position
and flags them to be ionized if they do.
After the final filtering step, pixels which have not 
yet been marked as fully ionized are allocated a 
value of $x_{\rm \textsc{hi}}=1-\zeta f_{\rm coll}(\bmath{r},z,R_{\rm pix})$, 
with the physical constraint that $0\le x_{\rm \textsc{hi}}\le 1$
to allow for partial ionization.

Once the neutral-fraction, density and differential 
velocity maps are 
calculated it is straightforward to 
calculate the brightness-temperature map 
by applying equation \ref{eqn:brightTemp} to every pixel. 
For a detailed description of this code we refer the reader 
to the {\small 21CMFAST} references quoted at the beginning of this section. 

\subsection{Local-outside-in: Pixel by pixel MHR00}

To construct outside-in models, in which under-dense regions
 are ionized first, we follow the general ideas
outlined by
\citet{Miralda-Escude2000} hereafter MHR00. 
The crucial concept in their model was that of a
 global recombination rate that is an increasing
 function of $\delta$. 
They imagine that all 
gas with a $\delta<\delta_{\rm ion}$ is ionized, whilst 
higher density gas is neutral due to this global 
recombination rate; $\delta_{\rm ion}$ is thus a threshold 
for ionization. Sources 
located in high-density regions, i.e. galaxies,
first ionize the dense gas of the galactic halo
after which the ionizing photons begin to
infiltrate into the IGM. Since
there are both less atoms to ionize and fewer recombinations
in low density regions, H\thinspace\textsc{ii} regions expand most 
quickly along directions of lowest gas density. 
The end of reionization is reached 
when ionized regions start overlapping and most 
of the IGM is illuminated by more than one source;
such overlap will occur in the lowest density 
`tunnels' that exist between sources.

To make a local version of the MHR00 model, we first 
smooth the density boxes generated from {\small 21CMFAST} with a 
k-space top-hat filter
on a radius of 0.75Mpc; this scale corresponds roughly to the 
Jeans length of $\sim$1Mpc for a gas temperature of
$\sim10^4$K \citep{Jeans2009}. The density box is 
numerical in nature, randomly generated from 
power spectrum initial conditions
and suffers from discretisation issues arising during 
the perturbation of mass that accounts for non-linear 
evolution;
smoothing counteracts these effects and makes the density maps more 
physical. 

We next establish the density threshold necessary 
to reproduce the volume-averaged neutral fraction 
$\overline{x}_{\rm \textsc{hi}}$ from our {\small 21CMFAST} simulation
when ionizing individual pixels whose overdensities satisfy
$\delta<\delta_{\rm ion}$. The threshold for ionization, 
$\delta_{\rm ion}$, is defined to be the overdensity in the $i^{\rm th}$
pixel that satisfies $i/N_{\rm pix} = 1-\overline{x}_{\rm \textsc{hi}}$
when the pixels are sorted into ascending order and 
$N_{\rm pix}$ is the total number of pixels. Finally, 
we create the $x_{\rm \textsc{hi}}$ map by ionizing all pixels that 
meet the condition $\delta<\delta_{\rm ion}$. 
Once we have our neutral-fraction maps we can generate the 
corresponding brightness-temperature maps. 

\subsection{Local-inside-out: pixel-by-pixel MHR00 inversion}
In creating a localised inside-out model we simply invert the
ionization condition of the local-outside-in model to $\delta>\delta'_{\rm ion}$.
Again, we fix the threshold to ensure that the 
box has the desired neutral fraction. This
is implemented with same method as the
local-outside-in model, except with
the density pixels in descending order. 
Once we have $\delta'_{\rm ion}$, we can
simply cycle through pixels and ionize any that
meet the above condition. The corresponding
brightness-temperature maps can then be calculated.

\subsection{Global-outside-in: Inverting {\small 21CMFAST}}

The framework of FZH04 can be used to enforce the 
prescription of MHR00 in a way that is sensitive to 
the large-scale nature of the density field. 
We suppose that over-dense
regions have too high a recombination rate 
for ionized regions to form and develop, 
instead a background of 
ionizing radiation of efficiency $\zeta_{\rm bg}$ 
ionizes under-dense regions of the IGM;
this efficiency factor can be parametrized
in a similar way to that of FZH04 but in the
context of more energetic radiation. 
We assume a critical density  $\delta''_{\rm ion}$ below which the 
recombination rate in a region is low enough for an ionized 
bubble to develop. 
We take the overdensity field
to be a function of position and smoothing scale ($R_{\rm smooth}$)
to consider $f_{\rm low}$, the fraction of 
matter on a given scale with 
$\delta(\bmath{r},R_{\rm smooth})\le\delta''_{\rm ion}$.
There is no clearly defined value that 
this critical density should take, so we choose 
$\delta''_{\rm ion}=-\delta_{\rm c}$ to aid in comparison
with our other models.
Given a region of mean overdensity $\delta_{m}$ 
the PDF 
is
$\exp{[ -(\delta-\delta_{m})^2/2(\sigma_{\rm min}^2-\sigma_m^2) ]}/\sqrt{2\pi(\sigma_{\rm min}^2-\sigma_m^2)}$, where we adopt the
same notation as in section \ref{sec:21cmfast}.
The fraction in a region that is sufficiently 
under-dense to potentially be ionized is then

\begin{equation}
\begin{split}
f_{\rm low} = \int_{-\infty}^{-\delta_{\rm c}}\d\delta\;
\frac{2}{\sqrt{2\pi(\sigma_{\rm min}^2-\sigma_m^2)}}\exp{\left[\frac{-(\delta-\delta_{m})^2}{2(\sigma_{\rm min}^2-\sigma_m^2)}\right]} 
\,.\\
\label{eqn:flow1}
\end{split}
\end{equation}
The factor of two arises from trajectories that reach the 
barrier but reflect, in this case, upwards \citep{Bond1991}. 
By making the substitution 
$x=(\delta-\delta_m)/\sqrt{2(\sigma_{\rm min}^2-\sigma_m^2)}$
we can write $f_{\rm low}=2/\sqrt{\pi}\int_{-\infty}^{x_{\rm c}}\exp{-x^2}\d x$,
where $x_{\rm c}=-(\delta_{\rm c}+\delta_m)/\sqrt{2(\sigma_{\rm min}^2-\sigma_m^2)}$,
which is refined as follows,
\begin{equation}
\begin{split}
f_{\rm low} 
&=\frac{2}{\sqrt{\pi}}\left[\int_{-\infty}^{0}\exp{(-x^2)}\d x+\int_{0}^{x_{\rm c}}\exp{(-x^2)}\d x\right]\\
&=1+\operatorname{erf}(x_{\rm c})=1-\operatorname{erf}(-x_{\rm c})\\
&=\operatorname{erfc}\left[ \frac{(\delta_{\rm c}+\delta_{m})}{\sqrt{2(\sigma_{\rm min}^2-\sigma_m^2)}}\right]
\,.\\
\label{eqn:flow2}
\end{split}
\end{equation}
\noindent 
The mass ionized in 
our underdensity is then related to the efficiency of the
ionizing photons by $m_{\rm ion}=\zeta_{\rm bg}m_{\rm low}$
so that

\begin{equation}
\begin{split}
f_{\rm low} &=  \frac{m_{\rm low}}{m_{\rm total}} = \zeta_{\rm bg}^{-1}\frac{m_{\rm ion}}{m_{\rm total}}
= \zeta_{\rm bg}^{-1} f_{\rm ion}\,;\\\label{eqn:ionCond}
\end{split}
\end{equation}
\noindent so for a region to be fully ionized we
require $f_{\rm low}\ge\zeta_{\rm bg}^{-1}$. This
translates into a barrier for ionization 
as follows,

\begin{equation}
\begin{split}
\operatorname{erfc}
\left[
\frac{(\delta_{\rm c}+\delta_{m})}
{\sqrt{2(\sigma_{\rm min}^2-\sigma_m^2)}}
\right]\
&\ge
\zeta_{\rm bg}^{-1}\,,\\
\operatorname{erf}^{-1}
\left(1-\zeta_{\rm bg}^{-1}\right)
&\ge
\frac{(\delta_{\rm c}+\delta_m)}
{\sqrt{2(\sigma_{\rm min}^2-\sigma_m^2)}}\,,\\
\sqrt{2(\sigma_{\rm min}^2-\sigma_m^2)}
\operatorname{erf}^{-1}
\left(1-\zeta_{\rm bg}^{-1}\right)
&\ge
\delta_{\rm c}+\delta_{m}\,,\\
\\ 
\end{split}
\end{equation}
\noindent and so our barrier is simply 
the inverse of the FZH04 barrier with 
the first down-crossing
corresponding to ionized regions,

\begin{equation}
\begin{split}
\delta_{m} 
&\le 
\sqrt{2(\sigma_{\rm min}^2-\sigma_m^2)}
K(\zeta_{\rm bg})-\delta_{\rm c} \,,\label{eqn:GlobMHRbarrier}
\end{split}
\end{equation}

\noindent where 
$K(\zeta_{\rm bg})=\operatorname{erf}^{-1}(1-\zeta_{\rm bg}^{-1})$.
This is applied numerically by smoothing the density
fields from large to small scales, at each smoothing scale
ionizing any pixel that meet the condition of equation
\ref{eqn:GlobMHRbarrier}. At the end, any pixels
that remain neutral are assigned a partial neutral
fraction according to $x_{\rm \textsc{hi}}=1-\zeta_{\rm bg}f_{\rm low}$
under the constraint $0\le x_{\rm \textsc{hi}}\le 1$.
For our simulations we choose that $\zeta_{\rm bg}$
take the same value as the efficiency parameter 
$\zeta$ of FZH04. This construction is 
of course somewhat artificial 
in that we choose our free variables 
to match those of FZH04 to aid model comparison 
and we emphasize that there is no physical basis 
for these choices. The 
key prediction of characteristic sizes from 
the FZH04 model is a 
product of the increasing barrier. Our barrier,
illustrated by
the dashed black line of Figure \ref{fig:barrier},
instead decreases with $\sigma^2(M)$; because 
it is first down-crossings with which we are
concerned, the effect is identical and a
different choice for the barrier height will 
not alter this.

Being a mirror of the standard FZH04 approach, 
one might expect the resulting neutral fraction to be identical. 
In reality there are differences in that the 
simulated density field includes non-linearities 
which have a different impact in an 
outside-in model, where the neutral 
field is correlated to the density field, 
to the anti-correlated inside-out model.
Furthermore, {\small 21CMFAST} directly tests the condition 
$f_{\rm coll}\ge\zeta_{\rm esc}$ rather than using the 
barrier; in doing so it calculates the analytic Press--Schechter 
$f_{\rm coll}$. Since the parametrically fitted \citet{Sheth1999} mass 
function of \citet{Jenkins2001} has better agreement to 
numerical simulations, a correction factor, $A^{\rm PS/J}$, 
is applied. This correction can be folded into the 
ionizing efficiency of our global-outside-in model,
but as it is not essential to generating large characteristic 
sized bubbles we choose to not include it. 
We thus find that the evolution 
of the mean neutral fraction 
with redshift, while close, is not identical to the other models.
For this reason and because the volume-averaged ionized 
fraction $\overline{x}_{\rm ion}=1-\overline{x}_{\rm \textsc{hi}}$ is an increasing 
function of time we present all results as functions 
of average ionized fraction.

\subsection{Simulation outline}\label{sec:simdesc}
The box size of all 
simulations is 300 Mpc on a side, initial conditions are sampled 
on to a high resolution 1200$^3$ pixel grid and all 
subsequent boxes are generated on a 600$^3$ pixel grid resolving 
0.5 Mpc. 
This simulation size and resolution is 
competitive and should be large enough to capture 
the important features of reionization cf. 
\citet{Harker2009, Bittner2011, Friedrich2011, Zaroubi2012, Iliev2013}.

The fiducial {\small 21CMFAST}, global-inside-out simulation
uses $\zeta=16$. This choice 
generates a best-case scenario in 
which evolution is strongest at the redshifts of 
optimum sensitivity for MWA and LOFAR, whilst ensuring 
that reionization is complete by $z\approx6$;
our results are therefore optimistic. 

We simulate from a redshift of 6 up to 14.75 in increments of 
$\Delta z = 0.25$. We analyse the full 3D boxes and 
so neglect any evolution with redshift that might occur over 
300 Mpc in the line of sight. We take the statistics 
of these constant redshift 3D simulated boxes as a proxy 
to the 2D slices at constant redshift that the various 
telescopes will observe. 

\section{Toy model for the evolution of the neutral field}\label{sec:toymodel}

Before delving into the statistics of our four simulated 
 models, we would like to 
consider a simple toy model for the evolution 
of the neutral-fraction 
field's PDF and its associated moments. 
In the limit that a discrete map be perfectly 
resolved, a pixel will either be totally neutral or entirely ionized.
The PDF for such a field, 
$P(x_{\rm \textsc{hi}}|\overline{x}_{\rm \textsc{hi}},\mathrm{resolved},\mathrm{discrete})$, 
will therefore consist of two weighted 
Delta-functions $\delta_{\rm D}$.
Making use of normalisation and the definition of 
$\overline{x}_{\rm \textsc{hi}}$, the PDF is found to be
\begin{equation}
\begin{split}
P(x_{\rm \textsc{hi}}|\overline{x}_{\rm \textsc{hi}}) = (1-\overline{x}_{\rm \textsc{hi}})\delta_{\rm D}(x_{\rm \textsc{hi}}) + \overline{x}_{\rm \textsc{hi}}\delta_{\rm D}(x_{\rm \textsc{hi}}-1)\,,\\
\end{split}
\end{equation}
\noindent where we have dropped the
discrete and resolved conditionals for brevity;
this implies that
\begin{equation}
\begin{split}
\langle x_{\rm \textsc{hi}}^n \rangle=\int_0^1P(x_{\rm \textsc{hi}}|\overline{x}_{\rm \textsc{hi}})x_{\rm \textsc{hi}}^n\d x_{\rm \textsc{hi}} = \overline{x}_{\rm \textsc{hi}} \,.\\
\end{split}
\end{equation}

Armed with this PDF we can calculate the evolution of the moments
of the neutral fraction as a function of the mean neutral fraction. 
Using angular brackets to denote volume averaging, we find 
for the variance $\sigma^2$, skew $S_3$ and kurtosis $K_4$ that
\begin{equation}
\begin{split}
\sigma^2
&=\langle x_{\rm \textsc{hi}}^2 \rangle-\overline{x}_{\rm \textsc{hi}}^2 = \overline{x}_{\rm \textsc{hi}}(1-\overline{x}_{\rm \textsc{hi}}) \,,\\
S_3
&=\langle(x_{\rm \textsc{hi}}-\overline{x}_{\rm \textsc{hi}})^3\rangle \\
&=\overline{x}_{\rm \textsc{hi}}(1-\overline{x}_{\rm \textsc{hi}})(1-2\overline{x}_{\rm \textsc{hi}}) \\\,
K_4
&=\langle(x_{\rm \textsc{hi}}-\overline{x}_{\rm \textsc{hi}})^4\rangle \\
&=\overline{x}_{\rm \textsc{hi}}(1-\overline{x}_{\rm \textsc{hi}})(3\overline{x}_{\rm \textsc{hi}}^2-3\overline{x}_{\rm \textsc{hi}}+1)\,.\\
\label{eq:anal_xHmoments}
\end{split}
\end{equation}

\noindent These statistics all tend to 
zero at $\overline{x}_{\rm \textsc{hi}}=0$ and $\overline{x}_{\rm \textsc{hi}}=1$ as they should, 
there is also symmetry under 
$\overline{x}_{\rm \textsc{hi}}\rightarrow 1-\overline{x}_{\rm \textsc{hi}}$.

\begin{figure}
  \centering
  \includegraphics{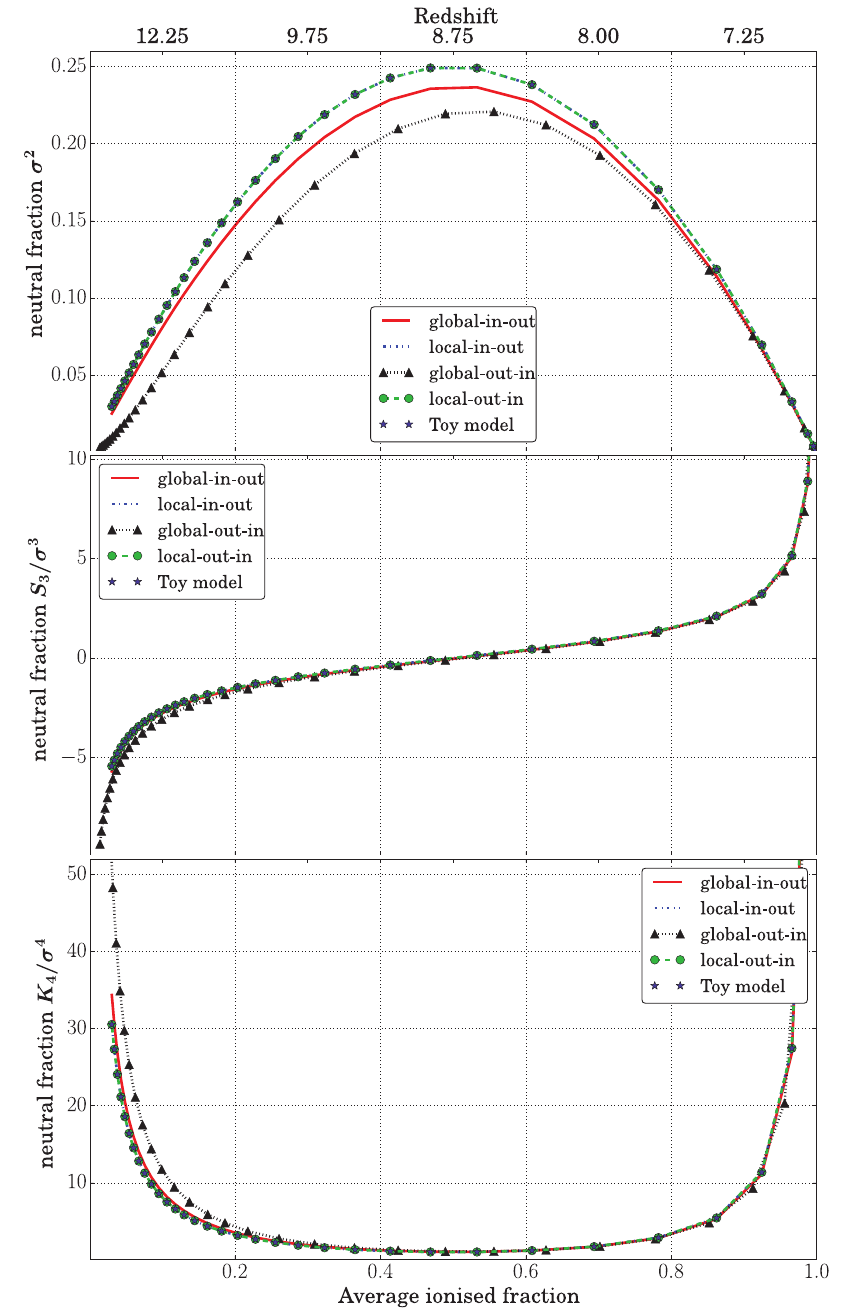}
  \caption{Theoretical prediction for moments of the 
neutral-fraction PDF
and those of the four simulations described in section 
\ref{sec:modeldesc}. From top to bottom 
variance, skewness 
and kurtosis of the $x_{\rm \textsc{hi}}$ field against 
$\overline{x}_{\rm ion}=1-\overline{x}_{\rm \textsc{hi}}$ are plotted. 
In the simulations the average of a box realisation 
is taken in calculating the central moments.
}
  \label{fig:xHmoments}
\end{figure}

We plot the analytic moments
normalised by $\sigma^n$ for the $n^{\rm th}$ moment along
with that of the $x_{\rm \textsc{hi}}$ 
maps from our four 
simulations against the average ionized fraction 
in Figure \ref{fig:xHmoments}. In this plot the curves correspond
to the toy model (purple stars), 
global-inside-out (red solid),
local-inside-out (blue dot-dashed),
global-outside-in (black dotted w/triangles), and
local-outside-in (green dashed w/circles). We will use the same 
model key in all figures hereafter.
We see 
that the local models identically
reproduce the distribution of the toy model.
This is because these models are constructed without partial ionizations
and are for all intents and purposes `perfectly resolved' $x_{\rm \textsc{hi}}$ maps  
(note that this is not true of the simulated density 
field used to generate them nor the
resulting brightness-temperature maps). 
Furthermore, it is only in the variance that we see 
any major deviation of the 
global models from that of the toy model. 

The variance is an inverted parabola where symmetry demands
that the maximum be at a mean ionized fraction of 0.5.
The skewness displays
strong evolution towards the end of reionization 
(where the average ionized fraction ranges
from 0.9 to 1); this feature translates 
on to the skewness of the brightness-temperature 
PDF and its
potential for use in detecting the end of reionization has been 
discussed previously \citep{Harker2009, Ichikawa2010}. 
The skewness passes from negative values to positive
at $\overline{x}_{\rm ion}=0.5$; we will see this also 
propagates onto the brightness-temperature skewness 
but only for inside-out models. 
The kurtosis shares the skewness' strong late-time 
evolution and also has a perfectly mirrored early-time signature
towards the beginning of reionization at mean ionized fraction of 
between 0 and 0.1. Unfortunately by this point the density 
field will dominate the statistics and this signature 
does not translate on to the brightness-temperature maps.
This is also the case for the inverted early-time 
singularity in the skewness, however with inside-out 
models this produces a detectable
early-time minimum as the brightness temperature becomes
increasingly influenced by the density field rather
than the neutral-fraction field with increasing redshift. 

\begin{figure}
  \centering
  \includegraphics{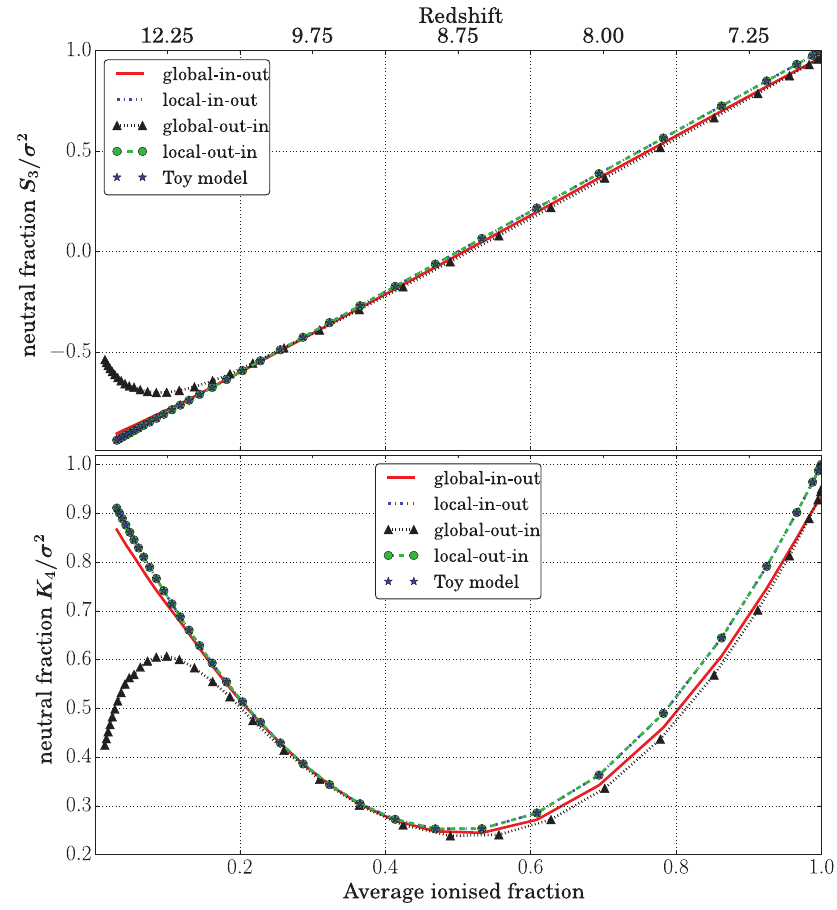}\\
  \caption{Theoretical prediction for dimensional higher 
order moments of the neutral-fraction PDF 
and those of the four simulations described in section 
\ref{sec:modeldesc}. From top to bottom we plot the 3rd and 4th 
central moments each normalised by $sigma^2$. 
}
  \label{fig:xHmoments2}
\end{figure}

Noting a common factor of $\sigma^2$ in all our analytic expressions
of equation \ref{eq:anal_xHmoments}, we plot the 
central moments of our toy model and simulated $x_{\rm \textsc{hi}}$ 
maps normalised by $\sigma^2$ in Figure \ref{fig:xHmoments2}. 
This appears to be a slightly more natural normalisation choice 
with both statistics remaining well behaved, 
exhibiting no singularities where $\sigma^2\rightarrow 0$ 
at either extreme of average ionized fraction. 
There is a strong deviation at small ionized fractions in 
the global-outside-in model 
caused by imperfect resolution coupled to a positive correlation 
with the density field. This same behaviour is observed in the 
global-inside-out model but to a lesser extent due to its anti-correlation 
with the density field.

We conclude that the central moments of the neutral-fraction 
PDF are essentially model independent, being
dominated by the PDF's bimodal nature and the distribution average. 
We therefore expect that any model differences discernible 
by measuring moments 
will result from the relationship between the $x_{\rm \textsc{hi}}$ 
and density fields. More importantly these 
signatures offer model independent flags of specific points 
of the reionization process, namely $\overline{x}_{\rm \textsc{hi}}>0.9$,
$\overline{x}_{\rm \textsc{hi}}=0.5$ 
and $\overline{x}_{\rm \textsc{hi}}<0.1$ 

\section{Clean brightness-temperature maps}\label{sec:CleanStats}

\begin{figure}
  \centering
  \includegraphics{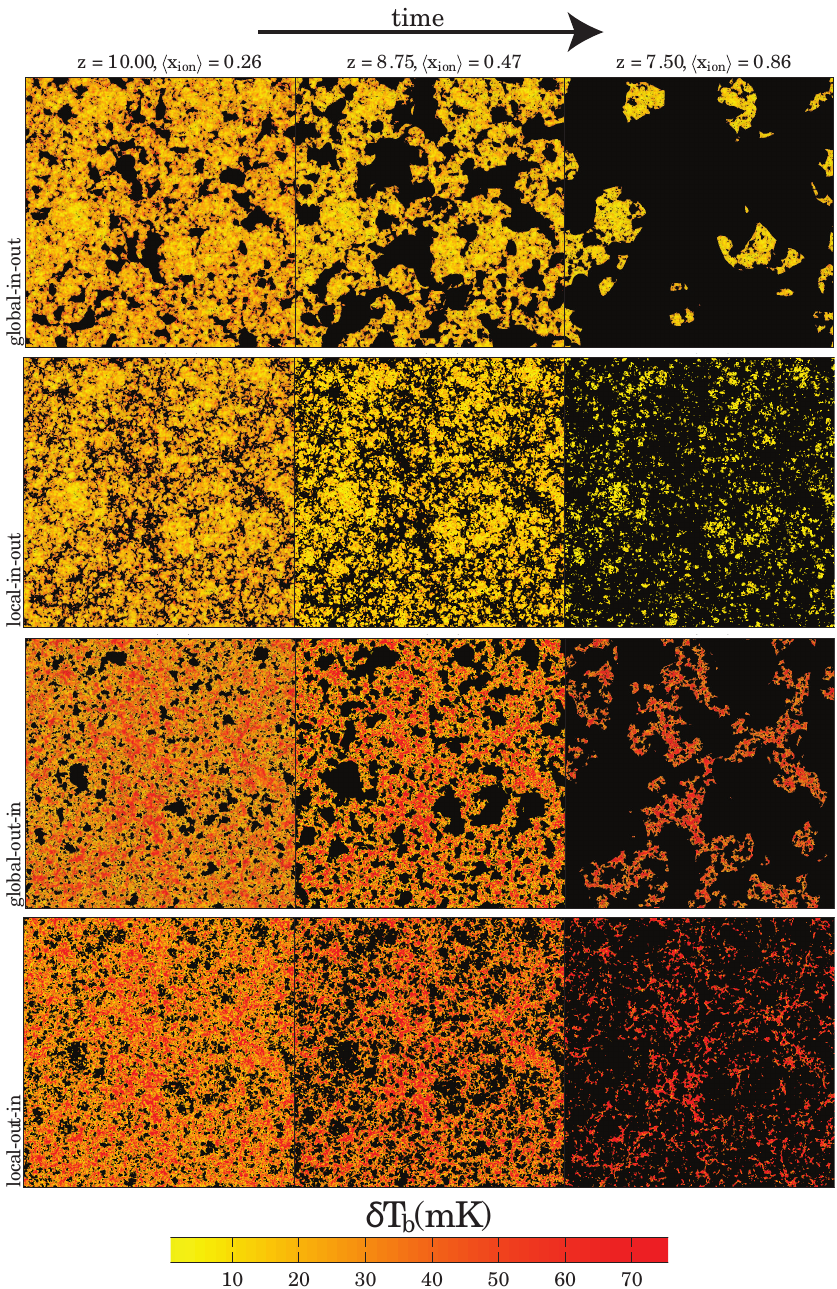} \\
  \caption{
Brightness-temperature maps for 300 Mpc box slices with
thickness of $0.5\,$Mpc.
Maps correspond to, from op to bottom:
global-in-out, local-in-out, global-out-in, and local-out-in.
Panels correspond to
$z=10.00$, $\overline{x}_{\rm ion}=0.26$ (left); 
$z=8.75$, $\overline{x}_{\rm ion}= 0.47$ (middle); \& $z=7.50$, $\overline{x}_{\rm ion}= 0.86$ (right) for all but global
outside-in which instead corresponds to, 
from left to right: $z=$ 9.75,
8.75, 7.50 and mean ionized fractions of 0.26, 0.49 and 0.85.
Black regions correspond to those that are fully ionized.
}
  \label{fig:dTSliceCompare}
\end{figure}

Before we consider instrumental noise it is informative 
to study the brightness-temperature maps from clean simulations. 
We begin by considering 2D maps from 
slices of each simulation;
Figure \ref{fig:dTSliceCompare} shows 
brightness-temperature maps for each, 
the depth of each slice is 0.5 Mpc. 
Maps
correspond from top to bottom to: global-in-out,
local-inside-out,
global-outside-in,
local-outside-in
and from left to right to $\overline{x}_{\rm ion}=0.26$, $0.47$ and $0.86$
except for global-outside-in which corresponds instead to 
$\overline{x}_{\rm ion}=0.26, 0.49$ and $0.85$. The black 
represents fully ionized regions in which there is 
no brightness-temperature signal. We see that 
at higher average ionized fractions the models appear very 
different. Both local models display a more randomised 
distribution of smaller islands of signal, with
local-outside-in possessing larger ionized regions 
than that of local-inside-out; this is due to the large voids
between the filaments being ionized preferentially
in the local-outside-in model. 
The global models look very different 
with large cohesive ionized regions dominating the 
maps. Global-outside-in has smaller cohesive 
neutral islands than the global-inside-out model, a 
result of the clustering of matter under gravity into 
a vast interconnected web.
This is easiest to see in
the $z=7.5$ global maps; in the outside-in model
the filaments and walls have remained neutral to produce
many connected neutral regions. In the inside-out 
the opposite is true and only the voids remain neutral, 
resulting in many connected ionized
regions around the filaments and walls. 
Concentrating efforts on imaging the later stages of reionization 
would be valuable, as at lower 
mean ionized fractions the ionized bubbles 
decrease in size and it becomes more difficult
to visually differentiate between models. 

\begin{figure}
  \centering
  \includegraphics{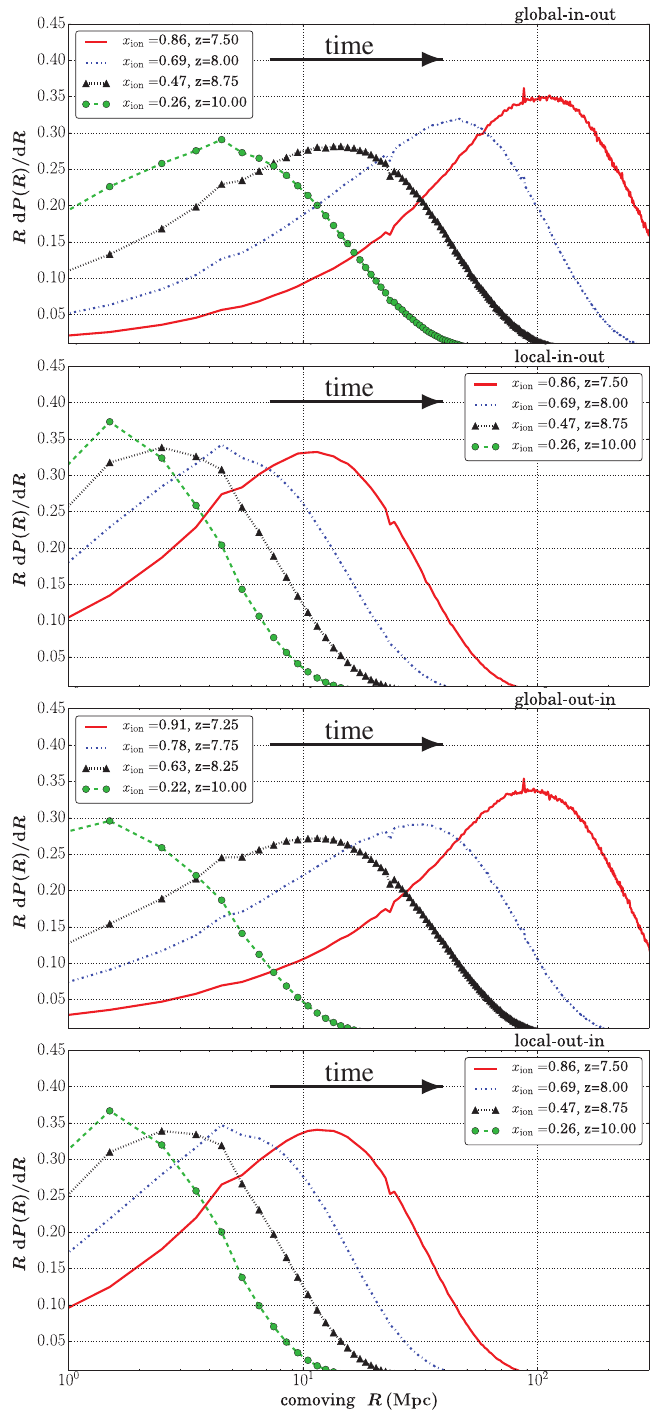}\\   
  \caption{Size distribution of ionized regions 
$R \d P(R)/\d R$, where $\d P(R)/\d R$ is the 
probability density function for regions of radius $R$ 
for the four models. Plots correspond to, from top to bottom: 
global-in-out, 
local-in-out, 
global-out-in,
 and local out-in. 
}
  \label{fig:SizeDist}
\end{figure}

To quantify this information we calculate the size distributions 
of the ionized regions in the brightness-temperature maps.
We calculate these using the Monte-Carlo method of \citet{Mesinger2007}
in which a random ionized pixel is selected, 
then the distance to the first ionized-to-neutral transition
recorded. This procedure is repeated $10^7$ times, from which 
the size distribution is approximated.
The global-outside-in model has slightly smaller bubbles than {\small 21CMFAST}
 for a given average ionized fraction. 
We therefore identify the average ionized fraction 
for which global-outside-in appears most like {\small 21CMFAST} 
in terms of bubble size. Size distributions 
for ionized regions are presented in figure 
\ref{fig:SizeDist}; from left to right curves correspond to 
average ionized 
fractions of 
0.26 (green dashed w/circles),
0.47 (black dotted w/triangles),
0.69 (blue dot-dashed), and
0.86 (red solid) for all but the global-outside-in model 
which instead corresponds to average ionized fractions of 
0.42 (green dashed w/circles),
0.63 (black dotted w/triangles),
0.78 (blue dot-dashed), and
0.91 (red solid). Both local models 
display very little evolution of size in comparison to the
global models. For example, ionized bubble sizes
are $< 80\,$Mpc in the local models
whereas for the global-inside-out they reach scales of
hundreds of Mpc. The global-outside-in develops
its larger bubbles later on in the process,
but as it is likely that we will not have an 
absolute measure of the average ionized fraction it will be 
difficult to distinguish these models by characteristic 
bubble size alone.
Put another way there is degeneracy between the inside-out/ outside-in 
nature of the model and the timing of reionization when
considering the characteristic size of the bubbles.

\begin{figure}
  \centering
  \includegraphics{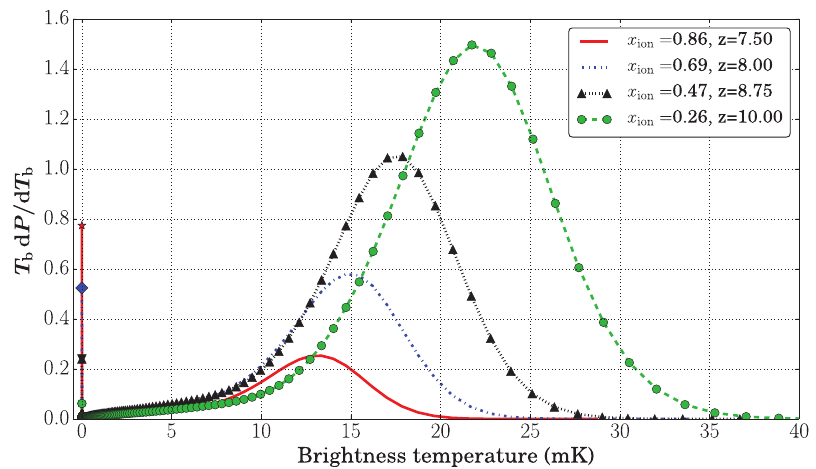}
  \caption{Brightness-temperature PDF from the 
global-inside-out simulation smoothed on a scale of 1.24 Mpc,
where $\mathrm{d}P/\mathrm{d}T_\mathrm{b}$ is the 
brightness temperature probability density.
Filled symbols at $\delta T_{\rm b}=0$
correspond to the fraction of fully ionized pixels 
to reduce the contrast between the delta function at $\delta T_{\rm b}=0$
and the non-zero distribution.} 
  \label{fig:PDF}
\end{figure}

Plotting the PDF of the brightness-temperature maps, 
shown in Figure \ref{fig:PDF} 
for the global-inside-out model, we see a
bimodal distribution as with the neutral-fraction field.
Curves correspond to 
$\overline{x}_{\rm ion}=0.26$ (green dashed w/circles),
$\overline{x}_{\rm ion}=0.47$ (black dotted w/triangles),
$\overline{x}_{\rm ion}=0.69$ (blue dot-dashed w/diamond) and 
$\overline{x}_{\rm ion}=0.86$ (red solid w/star). 
Note that
the filled symbols at $\delta T_{\rm b}=0$ correspond to 
the fraction of fully ionized pixels rather than their 
dimensionless probability density. 
We will not consider the kurtosis further here, 
because it will be more difficult to 
measure and behaves similarly to the 
skew, dominated by the interplay between the delta 
function at $x_{\rm \textsc{hi}}=0$ and a non-zero distribution.

\begin{figure}
  \centering
  \includegraphics{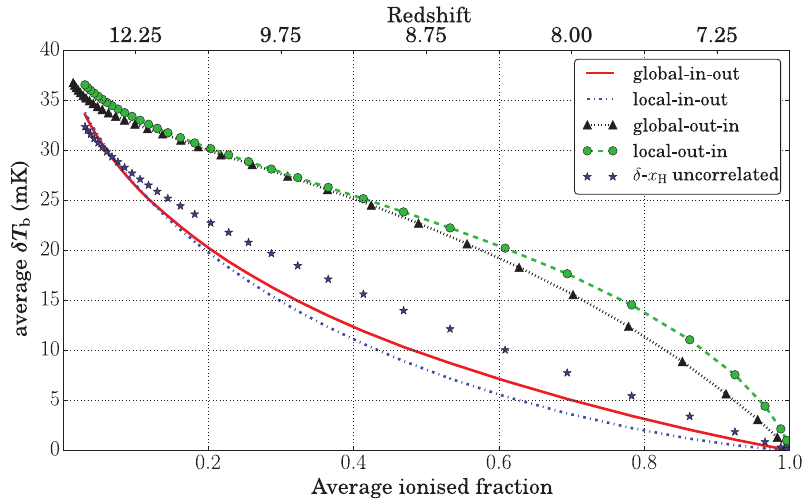}
  \caption{Evolution of average brightness temperature for the 
four models along with the predicted evolution assuming 
$\delta \overline{T}_{\rm b} \propto (1-\overline{x}_{\rm ion})$
(purple stars).
}
  \label{fig:nfevo&aveT}
\end{figure}

An important feature that
both outside-in models exhibit is considerably higher average temperatures 
than the inside-out models, where temperatures are 
boosted by the positive correlation 
between neutral-fraction and density field. Plotting 
the average temperature against average ionized 
fraction in Figure \ref{fig:nfevo&aveT}
illustrates this point; the outside-in models have
average temperatures nearly double that of the inside-out models for a 
large range of redshifts. From equation \ref{eqn:brightTemp} we see
that neglecting local velocities at a given redshift gives $\langle 
\delta T_{\rm b}\rangle \propto \langle x_{\rm \textsc{hi}}\rangle + \langle x_{\rm \textsc{hi}}\delta\rangle $, with $\langle x_{\rm \textsc{hi}}\delta\rangle $
boosting the signal 
when the neutral fraction and density are positively 
correlated and suppressing it when this 
correlation is negative.

As well as 21-cm fluctuation experiments such as
those considered here, 
the epoch of reionization may be constrained by 
observing the global or average 21-cm brightness 
temperature signal using single
dipole radio telescopes, e.g. Experiment to Detect the Global EoR
Signature (EDGES) \citep{bowman2010, Liu2013}.
We note that it would be prudent to account for the
inside-out/ outside-in nature of reionization when 
attempting to infer the neutral 
fraction from the observed average brightness temperature.
We see that even the fiducial model exhibits non-negligible
deviations from the assumption that 
$\delta \overline{T}_{\rm b} \propto (1-\overline{x}_{\rm ion})$,
plotted with purple stars in Figure \ref{fig:nfevo&aveT}, by
neglecting the anti-correlation between density and neutral-fraction fields.
Fortunately, as will be seen in section \ref{sec:NoisyStats},
establishing the inside-out/outside-in nature 
of reionization should not present a challenge.

\section{Statistics of `observed' brightness-temperature maps} \label{sec:NoisyStats}

Until this point we have considered the clean signal 
and what its statistics tell us about the underlying models.
However, this is far from the reality of actually observing these 
statistics. Any instrument will have a limited observing 
time ($t_{\rm int}$) and field of view 
($\rm FoV$); it will observe a range of scales
limited by both angular resolution ($\Delta\theta$) and 
frequency resolution ($\Delta\nu$). 
Finally, there is instrumental thermal 
noise to be considered and residuals left 
after the removal of bright radio foregrounds.
The usual approach to foreground removal is to
exploit the spectral smoothness of foregrounds; the
premise is that foregrounds can be fit, and therefore removed, 
using a low-order polynomial \citep{McQuinn2006,Wang2006}. 
To further reduce the effects of foregrounds it is 
possible to exclude from analysis frequency 
modes in which the foreground residuals are 
dominant; for example a `wedge' feature exists in the 
2D power spectrum in which the contribution from 
foreground residual noise is confined \citep{Bowman2009,Datta2010,Trott2012}.

In this work, we neglect foreground residuals and only 
consider instrumental effects. We take a simple approach 
to provide order of magnitude approximations to the noise 
errors on the various statistics we consider.
At the frequencies of relevance, it is reasonable to 
assume that the system temperature $T_{\rm sys}$ is saturated by
the sky temperature and that $T_{\rm sys}=180\,(\nu/180\rm{MHz})^{-2.6}$K 
in quiet portions of the sky \citep{Haslam1982}. 
The instrumental noise on the brightness temperature,
$\Delta T^N$,
measured by an interferometer is given by \citep{Furlanetto2006b}

\begin{equation}
\begin{split}
 \Delta T^N = \frac {T_{\rm sys}}{\eta_{\rm f}\sqrt{\Delta\nu t_{\rm int}} }\,.
\\
\label{eq:InstNoise1}
\end{split}
\end{equation}
\noindent Here the array filling factor is defined as 
$\eta_{\rm f}=A_{\rm tot}/D^2_{\rm max}$, where $A_{\rm tot}$ is 
the total effective area of the array and $D_{\rm max}$ is 
the maximum baseline of the array, controlling its resolution.

Throughout we present errors on the fiducial 
global-inside-out model. In the power spectrum and variance 
errors are comparable between the different models, but
the error on the skewness statistics is much less for
the outside-in models although they have a similar
qualitative evolution; as such errors on the skewness are pessimistic.

\subsection{Power spectrum of noisy maps}\label{sec:pserrors}
To date most attention has been focussed on measuring the
power spectrum. As such we will examine the evolution of the power spectrum 
for our four models before moving on to the moments. 
The power spectrum is the Fourier transform 
of the two-point correlation 
function which measures the level of correlation at 
different separations in real space. The two-point correlation 
is defined to be 
$\xi(\boldsymbol{r})=\langle \delta(\boldsymbol{x})\delta(\boldsymbol{x}+\boldsymbol{r})\rangle$
where the angle brackets denote an average over real space.

We concentrate on the dimensionless power spectrum which we describe using 
$\Delta^2_{\delta T_{\rm b}}(k,z)=k^3/(2\pi^2V)\langle |\delta_{21}(\bmath{k},z)|^2\rangle_k$ in which $\delta_{21}=\delta T_{\rm b}(\bmath{k},z)/\delta \overline{T}_{\rm b}(z)-1$, $\delta \overline{T}_{\rm b}(z)$ is the redshift dependent 
average brightness temperature calculated from the simulation, 
$V$ is the volume of the simulated
box and the angle brackets denote an average over $k$-space. 
By considering 
only the fluctuating variables in the differential 
brightness temperature, i.e. $\psi=x_{\rm \textsc{hi}}(1+\delta)$, 
\citet{Zaldarriaga2004} wrote the correlation function 
as,
\begin{equation}
\xi_{\psi}=\xi_{xx}(1+\xi_{\delta\delta})+\overline{x}_{\rm \textsc{hi}}^2\xi_{\delta\delta}+\xi_{x\delta}(2\overline{x}_{\rm \textsc{hi}}+\xi_{x\delta})\,.
\end{equation}

\noindent Here we see that the correlation function, and hence
the power spectrum, is dependent on the correlation 
of the neutral-fraction field $\zeta_{xx}$, the correlation 
function of the density field $\zeta_{\delta\delta}$ and the 
cross correlation of these two fields $\zeta_{x\delta}$.
This expression proves useful when considering 
the outside-in and inside-out models as their cross-correlation 
terms possess opposite signs.

\begin{figure}
  \centering
  \includegraphics{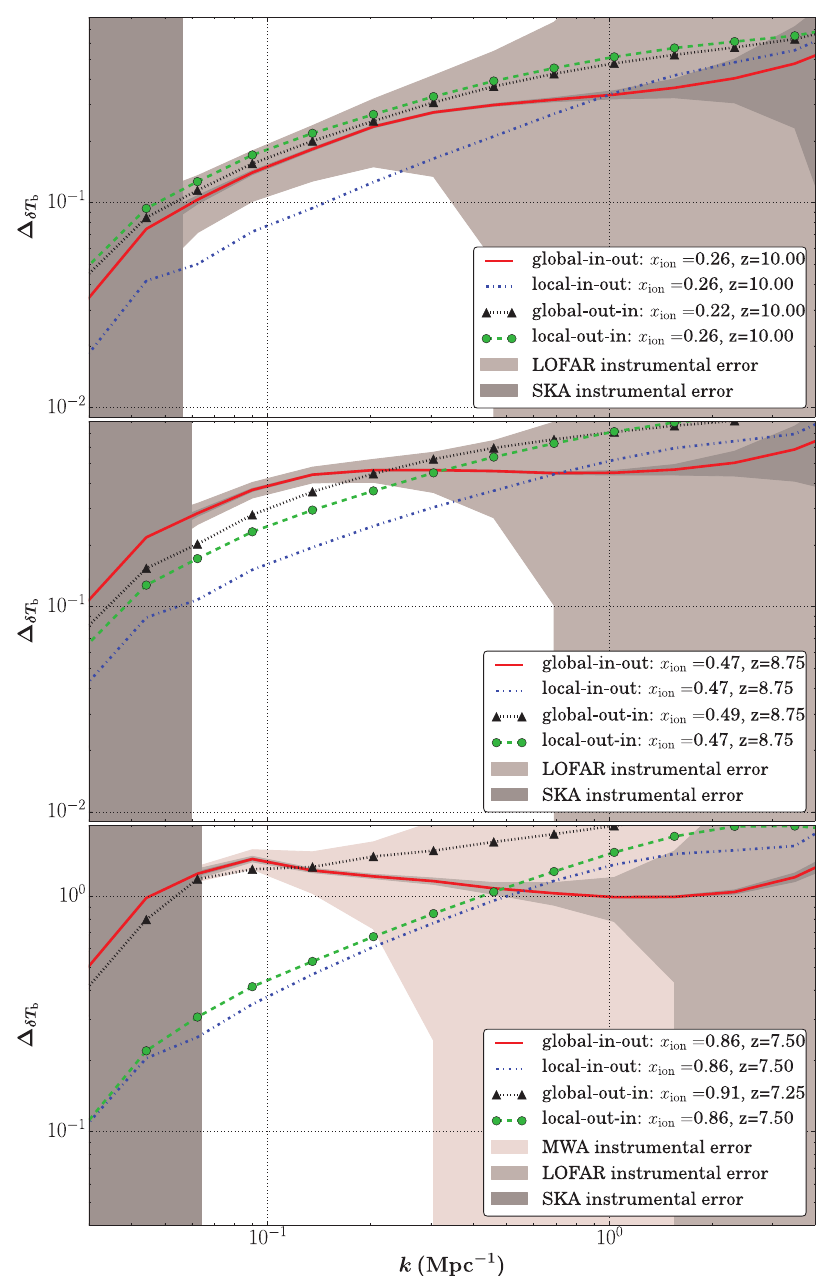}\\
  \caption{
Dimensionless brightness-temperature 
power spectrum with instrumental 
errors.
Plots correspond to $z=10.00$, $\overline{x}_{\rm ion}=0.26$ (top); $z=8.75$, $\overline{x}_{\rm ion}=0.47$ (middle); and $z=7.50$, $\overline{x}_{\rm ion}=0.86$ (bottom)
except for global-outside-in for which we plot
the redshift at which the power spectra are most similar. 
Beige shading depicts the 1-$\sigma$ instrumental errors 
for the global-inside-out model where
light to dark tones correspond to 
MWA, LOFAR and SKA respectively.
The cut off at small $k$ marks 
the largest scale that the instruments are sensitive
to, i.e. $2\pi/D_{\rm max}$. 
}
  \label{fig:pswerrors}
\end{figure}

To model instrumental errors on 
the spherically averaged power, we take the 
approach outlined in the appendix 
of \citet{McQuinn2006} adopting a logarithmic bin width of
$\epsilon=0.5$. We also assume uniform
UV coverage, of which none of the experiments we
consider can achieve, as such our power spectrum errors are
indicative only. 

Figure \ref{fig:pswerrors} shows
errors on the dimensionless brightness 
temperature power spectrum for the instrumental 
parameters outlined in Table \ref{tbl:instStats}. 
In this plot, and all those to follow, we use beige shading 
to depict the 1-$\sigma$ error on a quantity with tone from lightest to 
darkest corresponding to MWA, LOFAR and SKA respectively.
As is evident in this figure, the power spectrum of global-inside-out is 
strongly sensitive to the characteristic size, with a turnover 
in the spectrum providing a measure of its scale \citep{Signal2010}. 
\citet{Lidz2008} studied the observational repercussions
of this for MWA. They found 
it possible to constrain the amplitude and slope of 
the power spectrum for wave-numbers of between 
0.1-1$h$ Mpc$^{-1}$ and in this 
decade the amplitude rises and falls during 
reionization. The slope also
flattens out as reionization increases the size 
of the ionized bubbles. This results from a power 
boost at this characteristic size, 
which drops off on larger scales. 
As expected we see a very similar behaviour 
in the global-outside-in model, 
however the flattening out of the slope is less pronounced. 
This is because the positive correlation 
between $\delta$ and $x_{\rm \textsc{hi}}$ boosts 
the power most at smaller scales. 
By a mean ionized fraction of roughly 0.5, global-inside-out 
is the only model to display strong signatures
of a characteristic size in the power spectrum, as the additional 
power on small scales swamps out such a signature
in the global-outside-in model. 
Still, in the absence of a 
measure of the ionized fraction it will be difficult to 
know if the power spectrum is dropping rapidly 
with increasing redshift because reionization's 
procession was quick, or because 
the correct model is global-outside-in.
At earlier times/smaller average ionized fraction both 
outside-in models display more power than the
inside-out models due to the positive 
correlation of $x_{\rm \textsc{hi}}$ and the density field. 

If the process of
reionization follows a similar timing to that presented here,
LOFAR should be able to differentiate global models from
the small-scale boost in the global-outside-in model
towards the end of reionization,
but MWA will not. 
If the timing of reionization is less favourable and/or 
foreground residuals boost errors, even just by a bit, then 
LOFAR too could struggle to distinguish
the global models apart to any significance. 
However, SKA should not have a problem in
tightly constraining the models with just the 
power spectrum.

\begin{figure}
  \centering
  \includegraphics{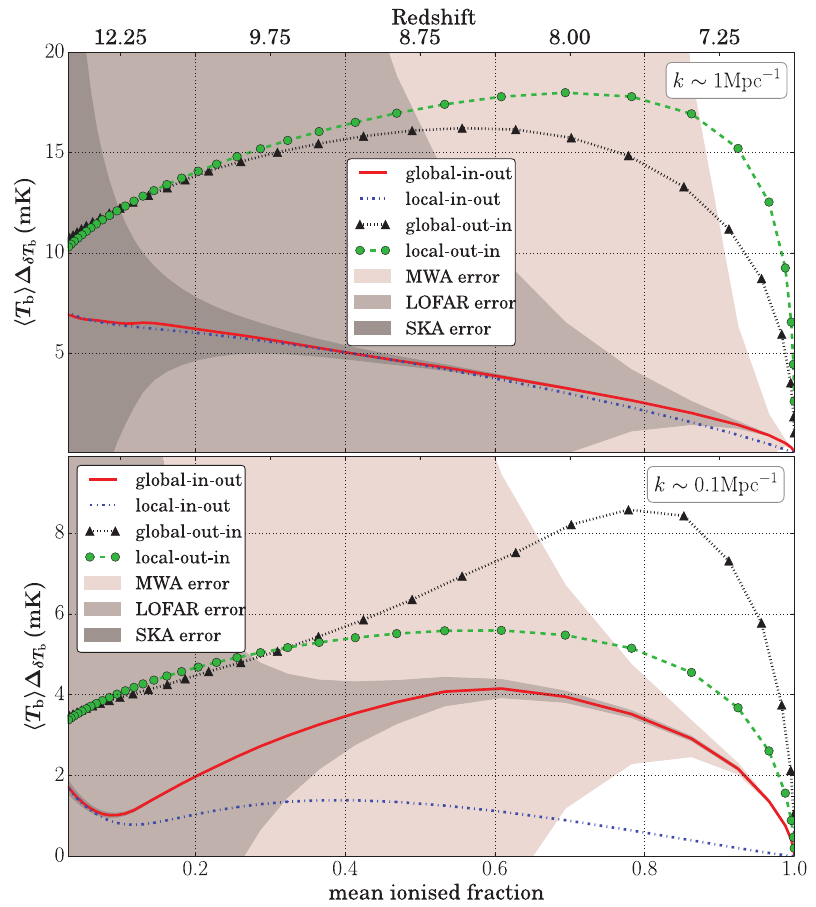}\\
  \caption{
Amplitude of the dimensional brightness-temperature 
power spectrum as a function of ionized fraction 
for $k=1\,$Mpc$^{-1}$ (top)
and $k=0.1\,$Mpc$^{-1}$ (bottom). 
Beige shading depict 1-$\sigma$ 
instrumental errors
for the global-inside-out model where
light to dark tones correspond to 
MWA, LOFAR and SKA respectively. 
}
  \label{fig:risenfall}
\end{figure} 

We plot the evolution of the dimensional power spectrum 
$\langle \delta T_{\rm b}\rangle\,\Delta(k)$ as a function of ionized fraction in
Figure \ref{fig:risenfall}. The first thing we notice 
is the higher amplitude of both outside-in models, 
which can be up to a factor of five greater in outside-in models
relative to inside-out models (ignoring comparison between 
global and local models). 
This property allows us to easily distinguish 
between inside-out and outside-in scenarios.
We also see that a dramatic rise and fall of the amplitude 
is common to both global models at $k\sim0.1\,$Mpc$^{-1}$, a feature
first noted for the global-inside-out model by \citet{Lidz2008}. 
Interestingly 
we see similar behaviour in both outside-in models for 
$k\sim1\,$Mpc$^{-1}$, which occurs because the 
positive correlation between neutral fraction
and small-scale overdensities
boosts the reionization mid-phase 
power, whilst the power is forced to that of the 
density field at early times and zero at late times.
Considering the instrumental errors 
in Figure \ref{fig:risenfall} we find that MWA should
distinguish inside-out from outside-in towards the end of reionization,
although not to great significance.
LOFAR and SKA should both be able to differentiate 
between outside-in and inside-out models with the 
dimensional power spectrum. 
Both LOFAR and SKA could also potentially 
constrain the timing of reionization using the 
global model's mid-phase maximum in the 
amplitude of the dimensional power spectrum at $k\sim 0.1\,$Mpc$^{-1}$
(or $k\sim 1\,$Mpc$^{-1}$ if global-outside-in transpires to be 
the correct model).

\subsection{Moments of noisy maps}

For the estimate of noise induced error on the moments we 
assume that there is independent noise in each pixel 
and that this noise is well described by a Gaussian random 
error with zero mean and standard deviation $\sigma_{\rm noise}$
derived from equation \ref{eq:InstNoise1} to be, 
\begin{equation}
\begin{split}
\sigma^2_{\rm noise}= 2.9 \rm{mK}&\left(\frac{10^5\rm{m}^2}{A_{\rm tot}}
                \right)\left(\frac{10'}{\Delta\theta}
                   \right)^2\\
                   &\times \left(\frac{1+z}{10.0}
                          \right)^{4.6}
                          \sqrt{\left(\frac{1\rm{MHz}}{\Delta\nu}\frac{100\rm{hours}}{t_{\rm int}}
                                     \right)} \,.
\\
\label{eq:InstNoise2}
\end{split}
\end{equation}

\noindent To derive this expression we have made use of the relation 
$D_{\rm max}=\lambda/\Delta\theta$.

\begin{table}
    \caption{Instrumental specifications assumed for noise calculations. 
LOFAR and SKA parameters are taken from \protect\citet{Mellema2013} and MWA 
parameters from \protect\citet{Tingay2013}.}
    \begin{center}
      \begin{tabular}[p]{||l||l||l||l||}
	\hline
        Parameter  &  MWA &  LOFAR &  SKA \\
        \hline
        Number of stations ($N_{\rm stat}$)
        & 128 & 48 & 450\\ 
        \hline
        Effective area ($A_{\rm eff}$/m$^2$)
        & 21.5 & 804&$10^6$/$N_{\rm stat}$\\ 
        \hline
        Maximum baseline ($D_{\rm max}$/m)
        & 2864 & 3000 & $10^4$\\ 
        \hline
        Integration time ($t_{\rm int}$/hours) 
        &1000 & 1000 & 1000 \\
        \hline
        Bandwidth ($B$/MHz) 
        &6 & 6 & 6 \\
	\hline
      \end{tabular}
    \end{center}
    \label{tbl:instStats} 
\end{table}

\noindent We assume the instrumental parameters of table
\ref{tbl:instStats} in working out the pixel noise properties. 
We fix the pixel size for the sake of computational 
efficiency to $R_{\rm pix}=6$ Mpc and $R_{\rm pix}=2$ Mpc 
to match the resolution of MWA/LOFAR and SKA
respectively.
We then proceed to smooth and re-sample the simulation boxes 
according to these pixel sizes and recalculate each statistic from these
`observed' boxes. 
We fix the angular resolution 
by the `observed' pixel size $R_{\rm pix}$ of our simulation.
These radio telescopes will have far better 
resolution in the frequency direction, but we bin to
improve signal to noise by matching the frequency resolution to 
simulation pixel size,
i.e. $\Delta\nu=H_0\nu_0\sqrt{\Omega_{\rm m}}R_{\rm pix}/[c \sqrt{(1+z)}]$.

The method for propagating the instrumental noise of 
equation \ref{eq:InstNoise2} on to errors on the moments are
detailed in the appendix. In short, we assume each 
pixel has a measured signal associated 
with it $x_i=\delta T_{i} + n_i$, where $\delta T_{i}$ 
is the true signal in the pixel and the noise on the pixel $n_i$ 
obeys the properties 
outlined at the start of this section. 
In this picture, we imagine that the simulation boxes 
represent some true measurable signal and we take central moments 
relative to the mean of the box 
$\delta \overline{T}_{\rm b} =N_{\rm pix}^{-1}\sum_{i=0}^{N_{\rm pix}} \delta T_{i}$. 
The true moments in which we are interested are
\begin{equation}
\begin{split}
S_2 &=  \frac{1}{N_{\rm pix}}\sum_{i=0}^{N_{\rm pix}}\left[\delta T_{i}-\delta\overline{T}_{\rm b}\right]^2\,,\\
S_3 &=  \frac{1}{N_{\rm pix}}\sum_{i=0}^{N_{\rm pix}}\left[\delta T_{i}-\delta\overline{T}_{\rm b}\right]^3\,,\\
K_4 &=  \frac{1}{N_{\rm pix}}\sum_{i=0}^{N_{\rm pix}}\left[\delta T_{i}-\delta\overline{T}_{\rm b}\right]^4\,,\\
\end{split}
\end{equation}
\noindent which correspond to the variance, skew and kurtosis 
respectively. 
We construct a test statistic for the $m^{\rm th}$ moment according
to $N_{\rm pix}^{-1}\sum_{i=0}^{N_{\rm pix}}(x_i-\overline{x}_i)^m$. 
Averaging this test statistic  over noise realisations,
we would expect to recover the 
true statistic if it was unbiased, any surplus is the bias of 
the test statistic and can
be used to construct an unbiased statistic. 
We found the skew test statistic to be unbiased, but 
that the naive test estimator for the variance does suffer from bias. 
This bias is removed by instead using
\begin{equation}
\begin{split}
\hat{S_2} &=  \frac{1}{N_{\rm pix}}\sum_{i=0}^{N_{\rm pix}}(x_i-\overline{x})^2 - \sigma_{\rm noise}^2 \,.\\
\end{split}
\end{equation}
\noindent The variance of each estimator
can then be calculated and 
propagated on to
the error on to $\gamma_3=\hat{S}_3/\hat{S}_2$ and $\gamma'_3=\hat{S}_3/(\hat{S}_2)^{3/2}$
to give

\begin{equation}
\begin{split}
V_{\hat{S}_2}&= 
\frac{2}{N} (2S_2\sigma_{\rm noise}^2 
+ \sigma_{\rm noise}^4)\,,\\
V_{\gamma_3} &\approx \frac{1}{(S_2)^2}V_{\hat{S}_3}
+\frac{(S_3)^2}{(S_2)^4}V_{\hat{S}_2}
-2\frac{S_3}{(S_2)^3}C_{S_2S_3} \,,\\
V_{\gamma'_3} &\approx \frac{1}{(S_2)^3}V_{\hat{S}_3}
+\frac{9}{4}\frac{(S_3)^2}{(S_2)^5}V_{\hat{S}_2}
-3\frac{S_3}{S_2^{4}}C_{\hat{S}_2\hat{S}_3} \,,\\
\end{split}
\end{equation}

\noindent where

\begin{equation}
\begin{split}
V_{\hat{S}_3}&= 
\frac{3}{N_{\rm pix}}(3\sigma_{\rm noise}^2K_4
+12S_2\sigma_{\rm noise}^4
+5\sigma_{\rm noise}^6)\,, 
\end{split}
\end{equation}
\noindent and
\begin{equation}
\begin{split}
C_{\hat{S}_2\hat{S}_3}
&= \frac{6}{N_{\rm pix}} 
S_3 \sigma^2_{\rm noise} 
\,.\\
\end{split}
\end{equation}

There is a further subtlety in that the value of $N$
corresponds to the number 
of pixels that the full FoV for each telescope would measure 
rather than the number of pixels in a simulated box. 
We must then work out the number of pixels that would fit into a 
single FoV. We assume a frequency depth or bandwidth $B=6$ MHz
over which the evolution in the brightness temperature is negligible;
the number of pixels measured is then taken to be 
$N=(L_{\rm FoV}/R_{\rm pix})^2(B/\Delta\nu)$, for which 
we approximate the FoV as a square of side $L_{\rm FoV}$. 

\subsection{Variance}

The variance is intimately related to the power spectrum;
it is the zero separation two--point correlation
function $\xi(0)$, which is 
equivalent to an integral over the power spectrum 
of fluctuations, i.e.
\begin{equation}
\begin{split}
\xi(\boldsymbol{r})=\int P(\boldsymbol{k})\exp{(i\boldsymbol{k}\cdot\boldsymbol{r})}\frac{d^3\boldsymbol{k}}{(2\pi)^3}\,,\\
\xi(\boldsymbol{0})=\sigma^2=\int P(\boldsymbol{k})\frac{d^3\boldsymbol{k}}{(2\pi)^3}\,.
\end{split}
\end{equation}

\begin{figure}
  \centering
  \includegraphics{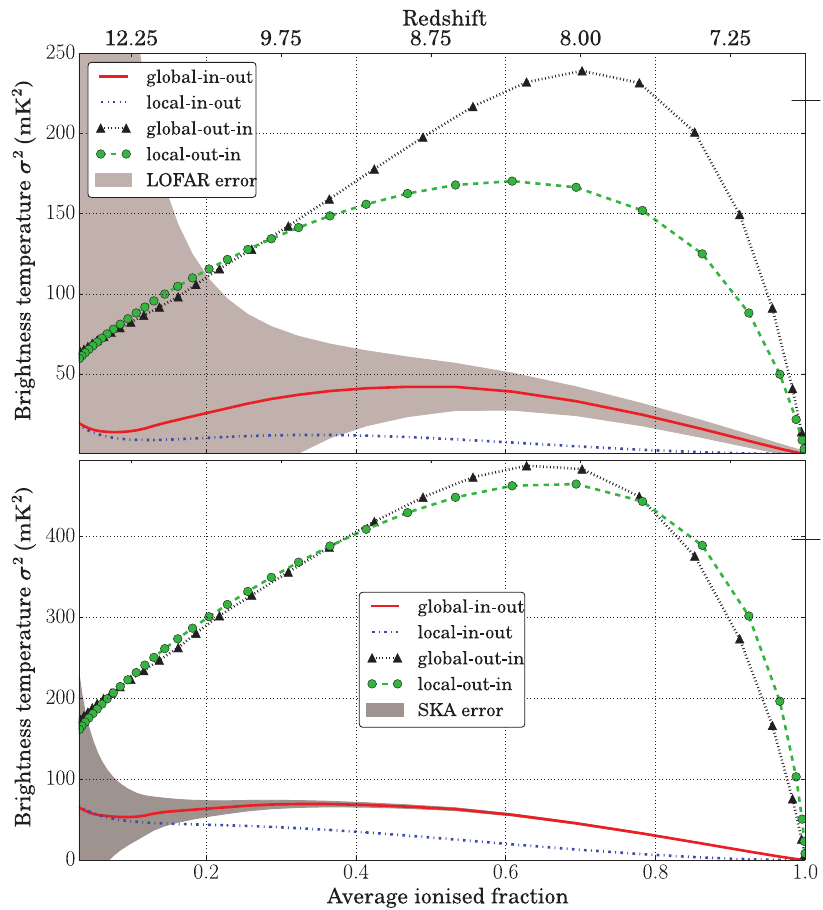}\\
  \caption{Variance of brightness temperature as measured in 
`observed' maps with a co-moving pixel size of 
6 Mpc for LOFAR (top) and 2 Mpc for SKA (bottom),
taken to be reflective of each instrument's resolution.   
Beige shading depicts 1-$\sigma$ 
instrumental errors 
for the global-inside-out model. 
}
  \label{fig:dTnoisyVariance}
\end{figure}

Whilst the variance does not offer 
any fundamentally new information,
the simplicity of this statistic will make 
it easier to measure and interpret. 
Furthermore instrumental 
effects are expected to affect statistics 
measured from maps differently than they will the power spectrum \citep{Petrovic2011}.
We calculate the variance in the $\delta T_{\rm b}$ 
maps according to

\begin{equation}
\begin{split}
\sigma^2=&\frac{1}{N_{\rm pix}}\sum_i^{N_{\rm pix}}\left[\delta T_{i}-\delta \overline{T}_{\rm b}\right]^2\,,\\
\label{eq:2ndmoment}
\end{split}
\end{equation}

\noindent here  $N_{\rm pix}$ is the total number of pixels in the
simulated box, 
$\delta T_{i}$ is the differential brightness 
temperature in the $i^{\rm th}$ pixel and 
$\delta \overline{T}_{\rm b}=N_{\rm pix}^{-1}\sum_i^{N_{\rm pix}}\delta T_{i}$ 
is the average in the simulated box. The evolution of this variance with 
average ionized fraction is shown in the bottom plot 
of Figure \ref{fig:dTnoisyVariance}.

We note that discretization effects of the density field induced  
when modelling of peculiar velocities were found to propagate
onto the brightness-temperature PDF, affecting the 
evolution of the moments; this effect is particularly
pronounced at early times when the 
density-field's statistics dominates the signal.
We find that resampling to a resolution of 
150 resolves the discretization issue, 
this produces pixels of side 2\,Mpc, 
the kind of resolution expected from SKA.
As a rule of thumb we find it sufficient to 
resample to one quarter of a simulation's
original resolution.

As expected, the variance shows little sensitivity to 
global or local nature, there are however 
strong differences in evolution between 
the inside-out models and the outside-in.
It is therefore extremely
useful for distinguishing between
these types of model.
These differences are due to the much
higher brightness temperatures of outside-in models
producing larger variance between the regions of
lowest or highest temperature and the map average. 
We conclude that 
observing very high variance 
would be indicative of an outside-in scenario.
The higher variance of outside-in models was 
qualitatively noted by \citet{Furlanetto2004} when
considering the brightness-temperature PDF 
during reionization.

If we consider the limits of this
evolution, we see that once reionization is complete there can
be no signal; at the other extreme, 
where it has yet to commence, 
the variance of the brightness temperature 
will be entirely defined by that of the density field.
As a result we observe an inverted parabola that 
peaks around the half way point for all models but local-inside-out. 
The peak is shifted right in the outside-in models where 
the variance 
is boosted until later because of the positive correlation 
between the density and the neutral-fraction fields
producing a more extreme non-zero distribution.

LOFAR, whose errors are shown in Figure 
\ref{fig:dTnoisyVariance} (top), should 
not be limited by instrumental noise 
in detecting the variance providing that nature has 
been kind with the timing of reionization. 
LOFAR will certainly be able to
exclude the outside-in models provided 
our observations that reionization was drawing 
to a close at $z\sim7$ are borne out. 
We find that the current planned 
configuration for MWA will not be 
able to overcome thermal noise in order to observe the variance
with resolution $\sim 6\,$Mpc. 
SKA on the other hand is expected to achieve small errors 
on the variance at most of the 
redshift range we have considered. As it is 
unlikely that reionization will have ended earlier than this, 
we conclude that there is great potential for distinguishing between 
models by measuring the variance with the SKA. 
We find all instruments to be better at constraining 
models and timing using the dimensional power spectrum 
rather than the variance at the default resolution we 
have considered in this section.

\subsection{Skewness}

As the variance is connected to the power spectrum
so too is the skew to the bi-spectrum.
The bi-spectrum is the Fourier transform of the 3-point 
correlation 
$\zeta(\boldsymbol{r_1},\boldsymbol{r_2})=\langle \delta(\boldsymbol{x})\delta(\boldsymbol{x}+\boldsymbol{r_1})\delta(\boldsymbol{x}+\boldsymbol{r_2})\rangle$,
where the angle brackets denote an average over real space 
or realisations; this measures the correlation between 3 
points in real space.
The skew is then equal to $\zeta(0,0)$, the zero separation 
3-point correlation function. The bi-spectrum is beyond
the scope of this paper but will be looked at in 
future work. Once reionization commences 
the brightness-temperature field becomes extremely non-Gaussian 
(as seen in the PDF of Figure \ref{fig:PDF}) 
and so it becomes necessary to consider higher order statistics 
than the power spectrum. The skew is therefore 
a very valuable statistic for teasing out information
beyond that contained in 2-point statistics, without 
the complexity of bi-spectrum measurements.

We consider the normalised third moment of the 
full brightness-temperature maps $S_3/\sigma^3$ (the skewness) where,
\begin{equation}
\begin{split}
S_3=&\frac{1}{N_{\rm pix}}\sum_i^{N_{\rm pix}}\left[\delta T_{i}-\delta \overline{T}_{\rm b}\right]^3 \,\,\,\,,\\
\label{eq:3rdmoment}
\end{split}
\end{equation}

\noindent $N_{\rm pix}$ is the total number 
of pixels in the map, $\delta T_{i}$ is
the differential brightness temperature of the $i^{\rm th}$ pixel and $\delta \overline{T}_{\rm b}$ is 
the average temperature in the box.

\begin{figure}
  \centering
  \includegraphics{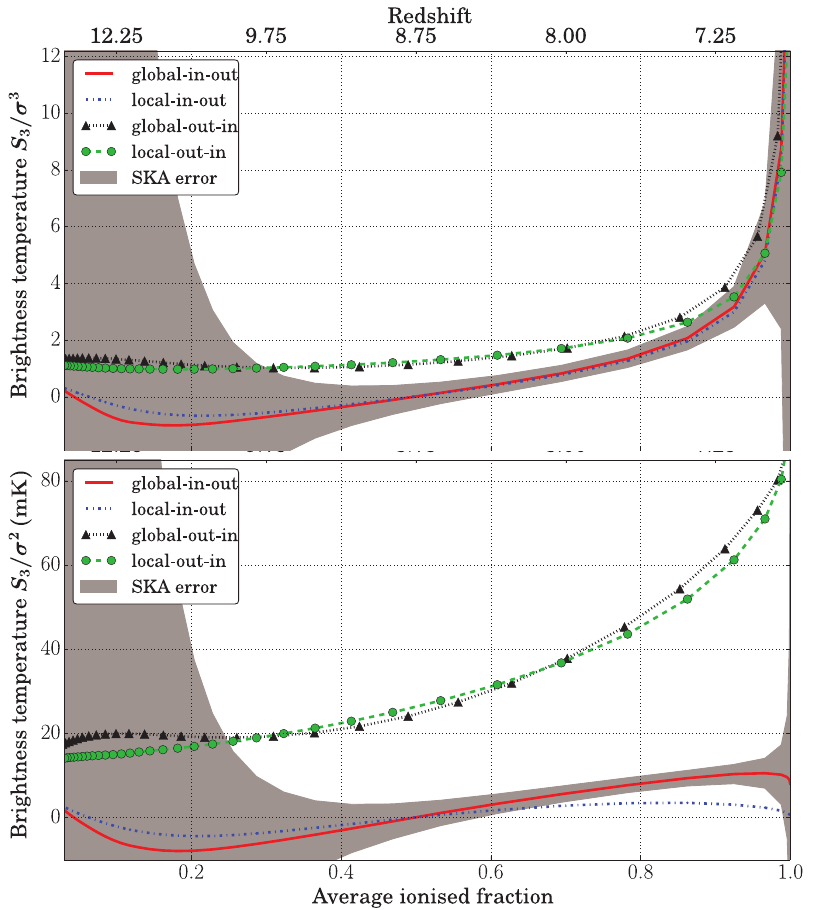}\\
  \caption{Skewness (top)
 and dimensional skewness (bottom) of the brightness 
temperature as measured in `observed' maps against 
average ionized fraction. 
Beige shading depicts 1-$\sigma$ 
instrumental errors
for the global-inside-out model.
Simulations are smoothed and re-sampled to reflect 
the default resolution of SKA, i.e. pixels of side 2 Mpc.
}
  \label{fig:dTnoisySkew}
\end{figure}

Figure \ref{fig:dTnoisySkew} (top) shows the skewness evolution for the four models.
As with the skewness of the neutral field we see a sharp 
increase at the end of reionization, i.e. $\overline{x}_{\rm ion}>0.9$;
again
this feature is found to be largely model independent, 
being strongly dominated by an
increasing delta function at $\delta T_{\rm b}=0$. 

As for distinguishing between models, we see differences up until 
$\overline{x}_{\rm ion}\sim 0.9$
These differences are slight, with the inside-out model's 
larger skewness originating in the 
positive $\delta$-$x_{\rm \textsc{hi}}$ correlation. 
Inspired by our analytical expressions for the 
moments of the neutral-fraction PDF we plot
the dimensional skewness $S_3/\sigma^2$
in Figure \ref{fig:dTnoisySkew} (bottom). 
We see that the inside-out and outside-in models
exhibit markedly different evolution of the dimensional
skewness. At early times inside-out models decrease with 
ionized fraction whereas outside-in increase; during the
mid-phases all display increasing dimensional skewness, then 
at later times the inside-out models drop off whilst the 
outside-in models increase in an almost exponential fashion.
It is useful to refer back to the dimensional skewness of
the neutral fraction (Figure \ref{fig:xHmoments2}),
we see that positive correlation between neutral fraction 
and density in the outside-in models serve to accentuate 
the features of the neutral-fraction field. 
In inside-out models this behaviour is ultimately suppressed as
high density regions are ionized over low density 
regions reducing the higher-temperature tail of 
the non-zero part of the PDF. 
Whilst strong model differences exist, we do already
have a means to constrain this behaviour in the variance. 
However, the dimensional 
skewness' model differences are very strong right up 
until the end of reionization, where as they are 
reduced in the  variance
as it is rapidly dropping off during this phase. 

The potential for the observing the skew on initial 
inspection seems less promising 
than for the variance; 
the noise dominates the skew entirely for LOFAR and MWA.
However for SKA the sharp increase 
signifying the end phase of reionization should not 
be noise dominated,
this is illustrated top plot of Figure \ref{fig:dTnoisySkew} 
which shows the evolution of skewness 
with 1-$\sigma$ SKA instrumental errors.
The suppression of variance in the maps
alters the evolution of the skewness. In
the inside-out model this results in an 
evolution much closer to that of the neutral-fraction maps,
being drawn to negative values as the skewness becomes
increasingly influenced by the neutral-fraction field
at early times such that a minimum is exhibited around
$\overline{x}_{\rm{ion}}\sim 0.2$, then 
during the mid-phase it passes from negative to positive. 
The evolution of the dimensional skewness 
$S_3/\sigma^2$ is presented in the bottom plot 
of Figure \ref{fig:dTnoisySkew}; suppression of the variance 
from smoothing the maps accentuates the features 
of this statistic. This is particularly 
valuable for identifying inside-out reionization which exhibits
strong negative values during the early stages of reionization
and passes from negative to positive at the mid-phase 
again echoing the evolution of the neutral-fraction's 
dimensional skewness.
This provides potentially useful 
signatures in the early minimum, mid-phase transition
from positive to negative, and late-time maximum in the 
global-inside-out's dimensional skewness.
Whilst the skewness already provides us with such a late-time
signature its similar early-time minimum is much weaker 
and may be harder to definitively detect.
However, from the 1-$\sigma$ errors in Figure \ref{fig:dTnoisySkew} 
it is clear that SKA will not be able to constrain 
the dimensional skewness very well with $2\,$Mpc pixels. 

\subsection{Smoothing `observed' maps to reduce noise}
It was noted by \citet{Harker2009} that smoothing their residual maps 
improved LOFAR's ability to constrain the skewness, something we find it
unable to do at a resolution of $6\,\rm Mpc$. We therefore
investigate the effect of further smoothing and re-sampling the 
`observed' brightness-temperature maps to increase pixel size 
and hence reduce pixel noise.

We find the model differences observed in the
brightness-temperature variance to be very robust 
to smoothing, although naturally its magnitude reduces 
with increased smoothing. 
As one might expect, smoothing most aggressively 
suppresses the variance at early and late 
times, when the field is Gaussian (early-times) or 
sparsely populated (late-times).
The mid-point maximum in inside-out models
remains robustly at $0.5 \le \overline{x}_{\rm ion}\le 0.6$ up to 
smoothing scales of $R_{\rm smooth}=60\,$Mpc, the largest smoothing
scale we consider. 
Although the local-inside-out's signal is smoothed 
out at much smaller smoothing scales than this 
as it has much lower variance than 
the other models; its early eradication under smoothing compared
to the other models is a signature of this. 
For global-outside-in, the mid-phase maximum 
shifts to higher $\overline{x}_{\rm ion}$ with increased smoothing
and so care would have to be taken in its 
interpretation under smoothing.
These differences in behaviour under increasing degrees 
of smoothing offer another method to differentiate 
between the four models we consider. 

In both global models the late-time increase 
in the skewness is robust up to smoothing scales 
of $R_{\rm smooth}\sim 50\,$Mpc beyond 
which it becomes too damped to contrast 
the $\overline{x}_{\rm ion}<0.1$ evolution sufficiently.
This signature is wiped out from the local models
at smoothing scales of $R_{\rm smooth}\sim 20\,$Mpc and beyond.
The early-time turnover in
the global-inside-out model is wiped out 
much beyond $R_{\rm smooth}\sim 20\,$Mpc, however it 
is robustly located at $\overline{x}_{\rm ion}\sim 0.2$
at lower smoothing scales.
Whilst the qualitative behaviour of the dimensional
skewness is robust to smoothing, the early and late-time
vertices become washed out beyond $R_{\rm smooth}\sim 15\,$Mpc.

\begin{figure}
  \centering
  \includegraphics{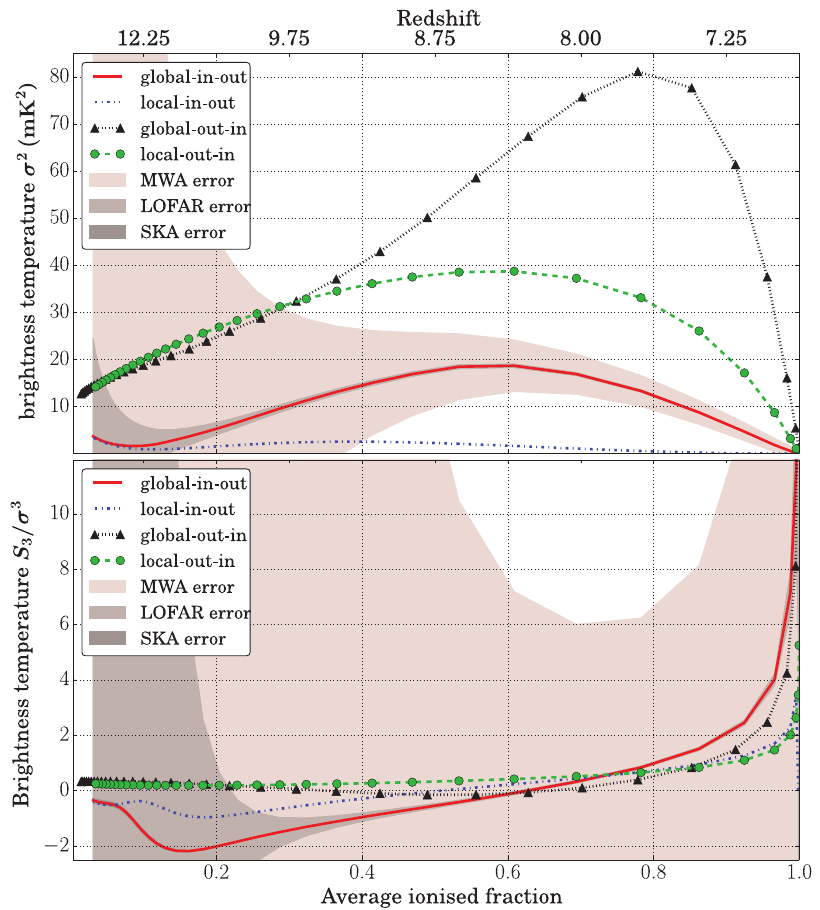}\\
  \caption{Brightness-temperature
variance (top) and skewness (bottom) as measured in `observed' maps 
smoothed on scales of 
$R_{\rm smooth}=10$ Mpc against 
average ionized fraction. 
Beige shading depicts 1-$\sigma$ 
instrumental errors 
on the global-inside-out model;
light to dark tones correspond 
to MWA, LOFAR and SKA errors respectively.
}
  \label{fig:smoothMom}
\end{figure}

We present the variance of `observed' maps
smoothed with a smoothing radius of $R_{\rm{smooth}}=10\,$Mpc 
in the top plot of Figure \ref{fig:smoothMom},
this illustrates that such smoothing allows 
LOFAR to constrain the variance
tightly out to $z\sim 12$ and renders SKA errors
negligible for the entire redshift range we consider. 
Whilst MWA's performance is not outstanding we
find that smoothing the maps up to $R_{\rm{smooth}}=30\,$Mpc 
or more allows MWA to better constrain the outside-in/ inside-out 
nature of our models using the 
variance than it can with 
the dimensional power spectrum; 
we find this to be the case even 
for smoothing of half this scale.

The bottom plot shows the skewness in maps again
smoothed to $R_{\rm{smooth}}=10\,$Mpc.
At this smoothing scale LOFAR will tightly 
constrain the skew up to $z\sim 10$ and SKA's 
errors are rendered negligible. 
Both LOFAR and SKA will be able to differentiate between models
using the skewness; in particular any detection of negative skewness
would be indicative of an inside-out model for reionization 
and $\overline{x}_{\rm ion}<0.6$ at the given redshift.
The mid-phase crossing from negative to 
positive will also be possible to constrain
although we note that it drifts to $\overline{x}_{\rm ion}\sim0.6$
with smoothing.
Furthermore, at this smoothing scale SKA is 
sensitive to the early-time $\overline{x}_{\rm ion}\sim0.2$
minimum in the skewness
of global-inside-out. 
Constraints of the early and mid-phase signatures will be 
even stronger using the dimensional skewness which 
has comparable errors to that of the skewness on this smoothing scale.
LOFAR will also be able to constrain this statistic 
until $z\sim 10$ at this smoothing scale, and in doing so
could constrain reionization's timing by the detection 
or absence of the early-time minimum and mid-phase transition
from negative to positive.
Again if the observed maps are smoothed to scales
of $R_{\rm{smooth}}=30\,$Mpc, MWA is able to constrain
the both skewness statistics up to $z\sim 9$.
It is worth noting that we also looked at the performance 
of the proposed instrument HERA \citep{Pober2013} and find its 
constraints on the variance to be almost as tight as those of 
SKA; HERA would also improve constraints on the skewness, pushing 
back sensitivity by $\Delta z\sim 2$ as compared to LOFAR constraints.

\section{Cosmic variance in noise-free maps} \label{sec:CosVar}

A box of 300 Mpc on a side is a large sample 
and we expect this to be representative of the 
Universe's total density field. However the process of 
reionization is extremely non Gaussian and so we expect 
measurements of the brightness temperature to suffer more 
extremely from cosmic variance, especially at the later 
stages when regions of neutral hydrogen will be 
increasingly rare. Another aspect of this is that
our simulation essentially follows one region of the 
Universe over a range of redshift; our measurements 
will come from independent regions of the Universe 
at different redshifts.
These different regions of the Universe will each have 
statistics that differ from each other and that 
of the full sample.
We attempt to gain some insight of this effect by using our 
simulation as a proxy for the full sample (the Universe)
and take sub-samples from our proxy to see how much each 
statistic varies. 
Of course, this gives us no insight into how
our simulation is biased from the full distribution; 
to estimate this would require repeat samples 
with different initial conditions, a less practical option 
given the scale of our simulation.
There is also the issue that our sub-samples are
smaller than our simulation box and so will suffer from a 
larger cosmic variance.
Even our simulation box size is different from the
instrument FoVs; for example the ratio of sub-sample volume
to instrumental volume is 0.003, 0.01 and 0.02 for MWA,
LOFAR and SKA respectively at a redshift of 8.5.
It is clear then that our estimate of cosmic variance 
will be larger than the true cosmic variance that the 
three experiments we consider will have to contend with.
As such we take our estimates as an upper-bound and 
use them for qualitative insight to the way the statistics
will be affected. 

We break our simulation box into 8 sub-samples 
of side 150 Mpc
and calculate the variance and skew statistics for each 
individual sub-sample. Each statistic's variance relative to 
that of the sub-sample population's mean 
is then calculated. We note that 
the sub-sample population mean of each statistic
is biased relative to that of the full-box due to 
the small sub-sample box size; this has been 
studied at length by \citet{Iliev2013} from which
we can conclude that our full box should not 
be biased in such a way.

\begin{figure}
  \centering
  \includegraphics{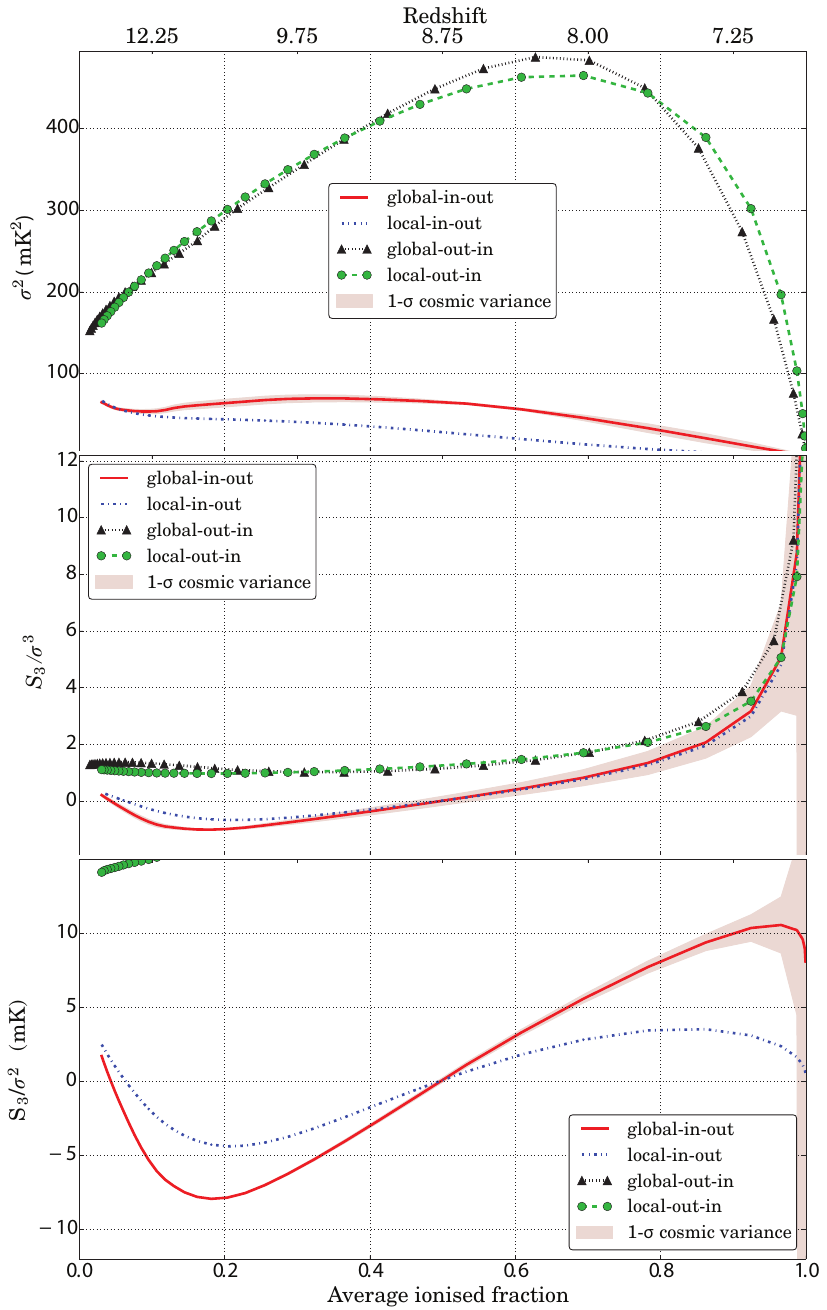}\\
  \caption{Cosmic variance induced errors as estimated from 
sub-sampling our full box into $2^3$ boxes of side 150 Mpc, then
calculating the sub-sample variance relative to the statistics of the full 
box for the brightness-temperature variance (top),
skewness (middle) and dimensional skewness (bottom).  
Beige shading depicts 1-$\sigma$ 
cosmic variance error
on the global-inside-out model.
}
  \label{fig:CosVar}
\end{figure}

The cosmic variance induced 1-$\sigma$ errors 
for the variance (top), skewness (middle) and dimensional 
skewness (bottom) are presented in Figure \ref{fig:CosVar}.
Cosmic variance should not pose any barrier in detecting 
the variance of the brightness temperature. 
If it transpires that global-outside-in
best describes the nature of reionization then the cosmic
variance error could be as big as 180$\,$mK during the mid-phases.
Of course the variance evolution is much more dramatic in outside-in 
models and so the conclusion that cosmic variance will not 
cause problems in measuring the variance holds. 
Cosmic variance will however impact  
detection of the skewness at late times, 
but it should be possible to detect an increase 
in skewness towards the end of reionization. 
The late-time maximum in the evolution of
dimensional skewness might not be evident, 
but the early-time equivalent in the inside-out 
models will be measurable. 
Combined with the late-time 
rapid increase in skewness, its mid-phase 
transition to negative
skewness and a corresponding maximum in the variance
we have detectable signatures of three distinct 
phases in the process of reionization.

\section{Conclusions} \label{sec:Conc}

In this paper a set of reionization
simulations have been built upon our reference model {\small 21CMFAST},
an efficient semi-numerical simulation that
is representative of our present understanding
of the likely nature of reionization.
Our four models are: {\small 21CMFAST},
a global-inside-out model in which large ionized bubbles
grow around large-scale overdensities;
global-outside-in, where large ionized bubbles
grow around large-scale underdensities;
local-inside-out, where small bubbles are
located in over-dense regions; and local-outside-in,
with small bubbles in under-dense regions.
We apply this testing set to the power
spectrum and the moments of the 21-cm brightness 
temperature 1-point distribution to tease
out how these statistics are influenced by the
general properties we model,
i.e. large or small bubbles located in under or
over-dense regions. We then consider the effect of instrumental noise
and discuss the practicalities of actually distinguishing
between models using observations from the three main 
radio telescopes that aim to measure the 21-cm fluctuations during
reionization, namely MWA, LOFAR and SKA. 
We then consider the challenge that 
the variability arising from only measuring limited 
regions of the Universe rather than the full 
Universe presents to interpreting observations 
with moments of brightness-temperature maps.

Outside-in models possess higher 
brightness temperatures than inside-out models 
due to the positive correlation
between density and neutral-fraction; this boosts
the variance and power spectrum accordingly.
The evolution of the average brightness temperature
in maps is found to be nearly double in outside-in
models; this must be accounted for when inferring
the neutral fraction from average brightness-temperature
measurements from global 21-cm surveys. We find
the variance to be the best option for distinguishing
the outside-in/inside-out nature of reionization; 
by smoothing noise out over 10$\,$Mpc or greater even
the pre-SKA telescopes will be able to achieve this 
out to redshifts of 12 (LOFAR) and 10 (MWA).
The detection of negative skewness is strongly
indicative of reionization being inside-out in nature,
the evolution of the dimensional skewness is also
very different in outside-in models, rising much
more aggressively during reionization. 
Again, with the help of smoothing, pre-SKA
instruments could exploit these differences
in out to redshifts of 10 (LOFAR) and 9 (MWA).

The moments primarily depend on the density to 
neutral-fraction correlation. As such it is only really 
the power spectrum that offers a means to constraining
the global/local nature of reionization. The power spectrum
show strong rise and fall in amplitude at scales 
of $k\sim0.1$ for both global models where power
is boosted by a large characteristic ionized bubble size.
It should be possible to use this signature, unique to global models, 
to exclude local models even with
pre-SKA instruments.
However, low sensitivity to the power spectrum may prevent
the pre-SKA instruments from distinguishing global-outside-in
models from global-inside-out, both of which exhibit 
a turnover at the scales of their characteristic 
bubbles. The degeneracy between
the power spectrum's evolution with redshift and the 
global-outside-in/inside-out nature complicates their 
separation further. SKA should be able
to use the power boost in outside-in models, 
evident at small scales late
in the reionization process and at all scales
earlier on, to distinguish these two models.

So long as reionization
does progress in a global-inside-out manner,
there will exist signatures of specific 
points of the reionization epoch in the variance
and skewness, all of which should be exploitable 
by pre-SKA instruments.
A minimum at $\overline{x}_{\rm ion}\sim0.2$
is evident in both skewness and dimensional skewness;
both then transition from negative to positive 
at $\overline{x}_{\rm ion}\sim0.5$ and the variance
reaches its maximum around this same time; finally
the skewness rises rapidly at $\overline{x}_{\rm ion}>0.9$.
Cosmic variance may make the late-time signature
in the skewness more of a challenge to detect, but
should not suppress its observation entirely.

There are of course several caveats to this work.
We present a single model for the evolution 
of the neutral fraction with redshift,
as such, 
reionization events described here are likely to be
shifted in redshift. This evolution was chosen
to present a best case scenario, our findings
for instrumental effects are therefore optimistic.
Our error treatment ignores correlations
in noise between pixels and therefore provides only
order of magnitude error estimates. 
Whilst such correlations are important, 
their effects are likely to be negligible in comparison to
error induced by foreground residuals,
which we make no attempt to incorporate in this work. 
Neglecting foregrounds allows for
clean interpretation of the underlying
statistics and understanding of instrumental effects.
We have only made a simple estimate of cosmic variance,
but this is sufficient to show it should not seriously
hamper efforts to constrain reionization with the 21-cm line.
Detailed modelling of
the spatial distribution of array stations
should be incorporated in future work,
but our modelling, whilst simplistic, 
is sufficient to estimate the errors.
Most of the characteristics we have considered
appear robust to smoothing so it is the hope
that this would also be the case with incomplete 
sampling; however, this should be investigated in full.
Finally, we have neglected spin temperature fluctuations;
if X-ray heating is very delayed compared 
to that considered here, then a large contrast
between a very cool neutral IGM and ionized regions
could produce a variance evolution comparable to that of the 
outside-in models \citep{Pritchard2007,Mesinger2013}. 

Four simple models for reionization have been explored,
showing that moments of the PDF show certain information
complementary to that of the power spectrum.
We find that pre-SKA instruments may struggle 
in distinguishing between global-inside-out and 
global-outside-in from the power spectrum alone 
but that they will be well placed
to break this degeneracy by measuring the variance.
Signatures of the early, mid and late phases of reionization
exhibited by the variance and skewness should be 
detectable at a wide range of redshift with the pre-SKA
instruments considered. The early and mid phase signatures
are more pronounced in the dimensional skewness, this 
statistic may therefore prove more robust to foreground residuals.
We conclude that whilst efforts to date have put
great emphasis on measuring the power spectrum, 
in the future, 
more focus must be placed on measuring the moments, 
whose simplicity make them much more accessible
in the short term.

\section*{Acknowledgements}
The authors would like to thank Daniel Mortlock and Andrew Jaffe for useful discussions and suggestions. We also thank Andrei Mesinger for making the {\small 21CMFAST} code used in this paper publicly available. CW is supported by an STFC studentship. JRP acknowledges support under FP7-PEOPLE-2012-CIG grant \#321933-21ALPHA and STFC consolidated grant ST/K001051/1.

\bibliographystyle{mn2e}

\begin{thebibliography}{}
 \providecommand{\href}[2]{#2}
  \providecommand{\doi}[1]{\href{http://dx.doi.org/#1}{doi:#1}}
  \providecommand{\eprint}[1]{\href{http://arxiv.org/abs/#1}{arXiv:#1}}

\bibitem[\protect\citeauthoryear{Alvarez \& Abel}{Alvarez \&
  Abel}{2012}]{Alvarez2012}
Alvarez M.~A.,  Abel T.,  2012, ApJ, 747, 126, \eprint{1003.6132}

\bibitem[\protect\citeauthoryear{Barkana \& Loeb}{Barkana \&
  Loeb}{2001}]{Barkana2001}
Barkana R.,  Loeb A.,  2001, Phys. Rep., 349, 125, \eprint{astro-ph/0010468v3}

\bibitem[\protect\citeauthoryear{Barkana \& Loeb}{Barkana \&
  Loeb}{2008}]{Barkana2008}
Barkana R.,  Loeb A.,  2008, MNRAS, 384, 1069, \eprint{0705.3246}

\bibitem[\protect\citeauthoryear{Becker et~al.,}{Becker
  et~al.}{2001}]{Becker2001}
Becker R.~H.  et~al., 2001, AJ, 122, 2850, \eprint{astro-ph/0108097}

\bibitem[\protect\citeauthoryear{Bennett et~al.,}{Bennett
  et~al.}{2013}]{Bennett2012}
Bennett C.~L.  et~al., 2013, ApJS, 208,
  20, \eprint{1221.5225}

\bibitem[\protect\citeauthoryear{Bittner \& Loeb}{Bittner \&
  Loeb}{2011}]{Bittner2011}
Bittner J.~M.,  Loeb A.,  2011, J. Cosmology \& Astroparticle Phys.,
  2011, 38, \eprint{1006.5460}

\bibitem[\protect\citeauthoryear{Bolton, Becker, Wyithe, Haehnelt \&
  Sargent}{Bolton et~al.}{2010}]{Bolton2010a}
Bolton J.~S.,  Becker G.~D.,  Wyithe J. S.~B.,  Haehnelt M.~G.,    Sargent W.
  L.~W.,  2010, MNRAS, 406, 612, \eprint{1001.3415}

\bibitem[\protect\citeauthoryear{Bolton, Haehnelt, Warren, Hewett, Mortlock,
  Venemans, McMahon \& Simpson}{Bolton et~al.}{2011}]{Bolton2011}
Bolton J.~S.,  Haehnelt M.~G.,  Warren S.~J.,  Hewett P.~C.,  Mortlock D.~J.,
  Venemans B.~P.,  McMahon R.~G.,    Simpson C.,  2011, MNRAS: Letters, 416, L70, \eprint{1106.6089}

\bibitem[\protect\citeauthoryear{Bond, Cole, Efstathiou \& Kaiser}{Bond
  et~al.}{1991}]{Bond1991}
Bond J.~R.,  Cole S.,  Efstathiou G.,    Kaiser N.,  1991, ApJ, 379, 440

\bibitem[\protect\citeauthoryear{Bowman \& Rogers}{Bowman \&
  Rogers}{2010}]{bowman2010}
Bowman J.~D.,  Rogers A.~E.,  2010, Nature, 468, 796

\bibitem[\protect\citeauthoryear{Bowman, Morales \& Hewitt}{Bowman
  et~al.}{2009}]{Bowman2009}
Bowman J.~D.,  Morales M.~F.,    Hewitt J.~N.,  2009, ApJ, 695, 183, \eprint{0807.3956}

\bibitem[\protect\citeauthoryear{Choudhury, Ferrara \& Gallerani}{Choudhury
  et~al.}{2008}]{Choudhury2008}
Choudhury T.~R.,  Ferrara A.,    Gallerani S.,  2008, MNRAS: Letters, 385, L58

\bibitem[\protect\citeauthoryear{Crociani, Mesinger, Moscardini \&
  Furlanetto}{Crociani et~al.}{2010}]{Crociani2010}
Crociani D.,  Mesinger A.,  Moscardini L.,    Furlanetto S.,  2010, MNRAS, 411, 13, \eprint{1008.0003}

\bibitem[\protect\citeauthoryear{Datta, Bowman \& Carilli}{Datta
  et~al.}{2010}]{Datta2010}
Datta A.,  Bowman J.~D.,    Carilli C.~L.,  2010, ApJ,
  724, 526, \eprint{1005.4071}

\bibitem[\protect\citeauthoryear{Dijkstra et 
al.}{2004}]{Dijkstra2004} Dijkstra M., Haiman Z., Rees M.~J., 
Weinberg D.~H., 2004, ApJ, 601, 666

\bibitem[\protect\citeauthoryear{Ewen \& Purcell}{Ewen \&
  Purcell}{1951}]{EWEN1951}
Ewen H.~I.,  Purcell E.~M.,  1951, Nature, 168, 356

\bibitem[\protect\citeauthoryear{Efstathiou}{1992}]
{Efstathiou1992} 
Efstathiou G., 1992, MNRAS, 256, 43P 

\bibitem[\protect\citeauthoryear{Fan, Narayanan, Strauss, White, Becker,
  Pentericci \& Rix}{Fan et~al.}{2002}]{Fan2002}
Fan X.,  Narayanan V.~K.,  Strauss M.~A.,  White R.~L.,  Becker R.~H.,
  Pentericci L.,    Rix H.-W.,  2002, AJ, 123, 1247, \eprint{astro-ph/0111184}

\bibitem[\protect\citeauthoryear{Fan, Strauss \& Becker}{Fan
  et~al.}{2006}]{Fan2006}
Fan X.,  Strauss M.~A.,    Becker R.~H.,  2006, AJ, 132,
  117, \eprint{astro-ph/0512082v2}

\bibitem[\protect\citeauthoryear{Field}{Field}{1958}]{Field1958}
Field G.~B.,  1958, Proc. IRE, 46, 240

\bibitem[\protect\citeauthoryear{Field}{Field}{1959}]{Field1959a}
Field G.~B.,  1959, ApJ, 129, 536

\bibitem[\protect\citeauthoryear{Friedrich, Mellema, Alvarez, Shapiro \&
  Iliev}{Friedrich et~al.}{2011}]{Friedrich2011}
Friedrich M.~M.,  Mellema G.,  Alvarez M.~A.,  Shapiro P.~R.,    Iliev I.~T.,
  2011, MNRAS, 413, 1353, \eprint{1006.2016v2}

\bibitem[\protect\citeauthoryear{Furlanetto \& Oh}{Furlanetto \& Oh}{2005}]{Furlanetto2005}
Furlanetto, S.~R., Oh, S.~P., 2005, MNRAS, 363, 1031, \eprint{astro-ph/0505065}

\bibitem[\protect\citeauthoryear{Furlanetto, {Peng Oh} \& Briggs}{Furlanetto
  et~al.}{2006}]{Furlanetto2006b}
Furlanetto S.~R.,  {Peng Oh} S.,    Briggs F.~H.,  2006, Phys. Rep., 433,
  181, \eprint{astro-ph/0608032v2}

\bibitem[\protect\citeauthoryear{Furlanetto, Zaldarriaga \&
  Hernquist}{Furlanetto et~al.}{2004}]{Furlanetto2004a}
Furlanetto S.~R.,  Zaldarriaga M.,    Hernquist L. [FZH04],  2004, ApJ, 613, 1, \eprint{astro-ph/0403697}

\bibitem[\protect\citeauthoryear{Furlanetto, Zaldarriaga \&
  Hernquist}{Furlanetto et~al.}{2004}]{Furlanetto2004}
Furlanetto S.~R.,  Zaldarriaga M.,    Hernquist L.,  2004, ApJ, 613, 16, \eprint{astro-ph/0404112v1}

\bibitem[\protect\citeauthoryear{Haiman, Abel \& Rees}{Haiman
  et~al.}{2000}]{Haiman2000}
Haiman Z.,  Abel T.,    Rees M.~J.,  2000, ApJ, 534, 11

\bibitem[\protect\citeauthoryear{Harker et~al.,}{Harker
  et~al.}{2009}]{Harker2009}
Harker G. J.~A.  et~al., 2009, MNRAS, 393, 1449, \eprint{0809.2428v2}

\bibitem[\protect\citeauthoryear{Haslam, Salter, Stoffel \& Wilson}{Haslam
  et~al.}{1982}]{Haslam1982}
Haslam C. G.~T.,  Salter C.~J.,  Stoffel H.,    Wilson W.~E.,  1982, A\&AS, 47

\bibitem[\protect\citeauthoryear{Heckman et~al.,}{Heckman
  et~al.}{2011}]{Heckman2011}
Heckman T.~M.  et~al., 2011, ApJ, 730, 5, \eprint{1101.4219}

\bibitem[\protect\citeauthoryear{Ichikawa, Barkana, Iliev, Mellema \&
  Shapiro}{Ichikawa et~al.}{2010}]{Ichikawa2010}
Ichikawa K.,  Barkana R.,  Iliev I.~T.,  Mellema G.,    Shapiro P.~R.,  2010,
  MNRAS, 406, 2521, \eprint{0907.2932}

\bibitem[\protect\citeauthoryear{Iliev, Mellema, Ahn, Shapiro, Mao \&
  Pen}{Iliev et~al.}{2013}]{Iliev2013}
Iliev I.~T.,  Mellema G.,  Ahn K.,  Shapiro P.~R.,  Mao Y.,    Pen U.-L.,
  2013, preprint, p.~20, \eprint{1310.7463}

\bibitem[\protect\citeauthoryear{Jeans}{Jeans}{1928}]{Jeans2009}
Jeans J.,  1928, {Astronomy and Cosmogony}.
Cambridge University Press

\bibitem[\protect\citeauthoryear{Jenkins, Frenk, White, Colberg, Cole, Evrard,
  Couchman \& Yoshida}{Jenkins et~al.}{2001}]{Jenkins2001}
Jenkins A.,  Frenk C.~S.,  White S. D.~M.,  Colberg J.~M.,  Cole S.,  Evrard
  A.~E.,  Couchman H. M.~P.,    Yoshida N.,  2001, MNRAS, 321, 372, \eprint{astro-ph/0005260}

\bibitem[\protect\citeauthoryear{Kim, Wyithe, Park \& Lacey}{Kim
  et~al.}{2013}]{Kim2013}
Kim H.-S.,  Wyithe J. S.~B.,  Park J.,    Lacey C.~G.,  2013, MNRAS, 433, 2476,  \eprint{1303.3051}

\bibitem[\protect\citeauthoryear{Lacey \& Cole}{Lacey \&
  Cole}{1993}]{LaceyCedric1993}
Lacey C.,  Cole S.,  1993, MNRAS,
  262, 627

\bibitem[\protect\citeauthoryear{Lidz, Zahn, Mcquinn, Zaldarriaga \&
  Hernquist}{Lidz et~al.}{2008}]{Lidz2008}
Lidz A.,  Zahn O.,  Mcquinn M.,  Zaldarriaga M.,    Hernquist L.,  2008, ApJ, 680, 962, \eprint{0711.4373v1}

\bibitem[\protect\citeauthoryear{Lidz, Faucher-Giguère, Dall’Aglio, McQuinn,
  Fechner, Zaldarriaga, Hernquist \& Dutta}{Lidz et~al.}{2010}]{Lidz2010}
Lidz A.,  Faucher-Giguère C.-A.,  Dall’Aglio A.,  McQuinn M.,  Fechner C.,
  Zaldarriaga M.,  Hernquist L.,    Dutta S.,  2010, ApJ, 718, 199, \eprint{0909.5210}

\bibitem[\protect\citeauthoryear{Liu, Pritchard, Tegmark \& Loeb}{Liu
  et~al.}{2013}]{Liu2013}
Liu A.,  Pritchard J.~R.,  Tegmark M.,    Loeb A.,  2013, Phys. Rev. D,
  87, 043002, \eprint{1005.4057}

\bibitem[\protect\citeauthoryear{Loeb \& Furlanetto}{Loeb \& Furlanetto}{2013}]{Loeb2013}
Loeb, A., Furlanetto, S.~R., 2013, {The First Galaxies in the Universe}, Princeton University Press

\bibitem[\protect\citeauthoryear{Madau, Meiksin \& Rees}{Madau
  et~al.}{1997}]{Madau1997}
Madau P.,  Meiksin A.,    Rees M.~J.,  1997, ApJ, 475,
  429, \eprint{astro-ph/9608010}

\bibitem[\protect\citeauthoryear{McQuinn, Zahn, Zaldarriaga, Hernquist \&
  Furlanetto}{McQuinn et~al.}{2006}]{McQuinn2006}
McQuinn M.,  Zahn O.,  Zaldarriaga M.,  Hernquist L.,    Furlanetto S.~R.,
  2006, ApJ, 653, 815, \eprint{astro-ph/0512263}

\bibitem[\protect\citeauthoryear{Mellema, Iliev, Pen \& Shapiro}{Mellema
  et~al.}{2006}]{Mellema2006}
Mellema G.,  Iliev I.~T.,  Pen U.-L.,    Shapiro P.~R.,  2006, MNRAS, 372, 679

\bibitem[\protect\citeauthoryear{Mellema et~al.,}{Mellema
  et~al.}{2013}]{Mellema2013}
Mellema G.  et~al., 2013, Exp. Astron., 36, 235, \eprint{1210.0197}

\bibitem[\protect\citeauthoryear{Mesinger, Ewall-Wice \& Hewitt}{Mesinger, Ewall-Wice \& Hewitt}{2013}]{Mesinger2013}
Mesinger, A., Ewall-Wice, A. \& Hewitt, J., 2013, preprint, \eprint{1311.4574}

\bibitem[\protect\citeauthoryear{Mesinger \& Furlanetto}{Mesinger \&
  Furlanetto}{2007}]{Mesinger2007}
Mesinger A.,  Furlanetto S.~R.,  2007, ApJ, 669, 663, \eprint{0704.0946v1}

\bibitem[\protect\citeauthoryear{Mesinger, Furlanetto \& Cen}{Mesinger
  et~al.}{2011}]{Signal2010}
Mesinger A.,  Furlanetto S.~R.,    Cen R.,  2011, MNRAS, 411, 955, \eprint{1003.3878v1}

\bibitem[\protect\citeauthoryear{Miralda-Escude}{Miralda-Escude}{2003}]{MiraldaEscude2003}
Miralda-Escude J.,  2003, ApJ, 597, 66, \eprint{astro-ph/0211071}

\bibitem[\protect\citeauthoryear{Miralda-Escude, Haehnelt \&
  Rees}{Miralda-Escude et~al.}{2000}]{Miralda-Escude2000}
Miralda-Escude J.,  Haehnelt M.,    Rees M.~J. [MHR00],  2000, ApJ, 530, 1-16, \eprint{astro-ph/9812306}

\bibitem[\protect\citeauthoryear{Morales \& Wyithe}{Morales \&
  Wyithe}{2010}]{Morales2010}
Morales M.~F.,  Wyithe J. S.~B.,  2010, ARA\&A, 48, 127, \eprint{0910.3010}

\bibitem[\protect\citeauthoryear{Mortlock et~al.,}{Mortlock
  et~al.}{2011}]{Mortlock2011}
Mortlock D.~J.  et~al., 2011, Nature, 474, 616, \eprint{1106.6088}

\bibitem[\protect\citeauthoryear{Ono et~al.,}{Ono et~al.}{2012}]{Ono2012}
Ono Y.  et~al., 2012, ApJ, 744, 83, \eprint{1107.3159}

\bibitem[\protect\citeauthoryear{Ota et~al.,}{Ota et~al.}{2010}]{Ota2010}
Ota K.  et~al., 2010, ApJ, 722, 803, \eprint{1008.4842}

\bibitem[\protect\citeauthoryear{Pan \& Barkana}{Pan \&
  Barkana}{2012}]{Pan2012}
Pan T.,  Barkana R.,  2012, preprint, \eprint{1209.5751}

\bibitem[\protect\citeauthoryear{Pentericci et~al.,}{Pentericci
  et~al.}{2011}]{Pentericci2011}
Pentericci L.  et~al., 2011, ApJ, 743, 132, \eprint{1107.1376}

\bibitem[\protect\citeauthoryear{Petrovic \& Oh}{Petrovic \&
  Oh}{2011}]{Petrovic2011}
Petrovic N.,  Oh S.~P.,  2011, MNRAS, 413, 2103, \eprint{1010.4109v2}

\bibitem[\protect\citeauthoryear{Planck Collaboration}{Planck Collaboration}{2013}]{PlanckCollaboration2013b}
Planck Collaboration: Ade P. A.~R.  et~al., 2013, preprint, \eprint{1303.5076}

\bibitem[\protect\citeauthoryear{Pober et~al.,}{Pober et~al.,}{2013}]{Pober2013}
Pober J.~C. et~al., 2013, preprint, \eprint{1310.7031}

\bibitem[\protect\citeauthoryear{Pritchard \& Furlanetto}{Pritchard \& Furlanetto}{2007}]{Pritchard2007} Pritchard, J.~R., Furlanetto, S.~R., 2007. MNRAS, 376, 1680, \eprint{astro-ph/0607234}

\bibitem[\protect\citeauthoryear{Pritchard \& Loeb}{Pritchard \&
  Loeb}{2008}]{Pritchard2008}
Pritchard J.~R.,  Loeb A.,  2008, Phys. Rev. D, 78, \eprint{0802.2102}

\bibitem[\protect\citeauthoryear{Pritchard \& Loeb}{Pritchard \&
  Loeb}{2011}]{Pritchard2011}
Pritchard J.~R.,  Loeb A.,  2011, Progress Phys., p.~64, \eprint{1109.6012}

\bibitem[\protect\citeauthoryear{Prodell \& Kusch}{Prodell \&
  Kusch}{1952}]{Prodell1952}
Prodell A.,  Kusch P.,  1952, Phys. Rev., 88, 184

\bibitem[\protect\citeauthoryear{Robertson et~al.,}{Robertson
  et~al.}{2013}]{Robertson2013}
Robertson B.~E.  et~al., 2013, ApJ, 768, 71, \eprint{1301.1228}

\bibitem[\protect\citeauthoryear{Schenker, Stark, Ellis, Robertson, Dunlop,
  McLure, Kneib \& Richard}{Schenker et~al.}{2012}]{Schenker2012}
Schenker M.~A.,  Stark D.~P.,  Ellis R.~S.,  Robertson B.~E.,  Dunlop J.~S.,
  McLure R.~J.,  Kneib J.-P.,    Richard J.,  2012, ApJ,
  744, 179, \eprint{1107.1261v2}

\bibitem[\protect\citeauthoryear{Sheth \& Tormen}{Sheth \&
  Tormen}{1999}]{Sheth1999}
Sheth R.~K.,  Tormen G.,  1999, MNRAS, 308, 119, \eprint{astro-ph/9901122}

\bibitem[\protect\citeauthoryear{Storrie-Lombardi, McMahon, Irwin \&
  Hazard}{Storrie-Lombardi et~al.}{1994}]{Storrie-Lombardi1994}
Storrie-Lombardi L.~J.,  McMahon R.~G.,  Irwin M.~J.,    Hazard C.,  1994, ApJ, 427, L13

\bibitem[\protect\citeauthoryear{Theuns, Schaye, Zaroubi, Kim, Tzanavaris \&
  Carswell}{Theuns et~al.}{2002}]{Theuns2002}
Theuns T.,  Schaye J.,  Zaroubi S.,  Kim T.-S.,  Tzanavaris P.,    Carswell B.,
   2002, ApJ, 567, L103, \eprint{astro-ph/0201514}

\bibitem[\protect\citeauthoryear{Tingay et~al.,}{Tingay
  et~al.}{2013}]{Tingay2013}
Tingay S.~J.  et~al., 2013, Publ. Astron. Soc.
  Australia, 30, 1, \eprint{1206.6945}

\bibitem[\protect\citeauthoryear{Trott, Wayth \& Tingay}{Trott
  et~al.}{2012}]{Trott2012}
Trott C.~M.,  Wayth R.~B.,    Tingay S.~J.,  2012, ApJ,
  757, 101, \eprint{arXiv:1208.0646v1}

\bibitem[\protect\citeauthoryear{Wang, Tegmark, Santos \& Knox}{Wang
  et~al.}{2006}]{Wang2006}
Wang X.,  Tegmark M.,  Santos M.~G.,    Knox L.,  2006, ApJ, 650, 529, \eprint{astro-ph/0501081}

\bibitem[\protect\citeauthoryear{Wyithe \& Morales}{Wyithe \&
  Morales}{2007}]{Wyithe2007a}
Wyithe J. S.~B.,  Morales M.~F.,  2007, MNRAS, 379, 1647, \eprint{astro-ph/0703070}

\bibitem[\protect\citeauthoryear{Zaldarriaga, Furlanetto \&
  Hernquist}{Zaldarriaga et~al.}{2004}]{Zaldarriaga2004}
Zaldarriaga M.,  Furlanetto S.~R.,    Hernquist L.,  2004, ApJ, 608, 622, \eprint{astro-ph/0311514}

\bibitem[\protect\citeauthoryear{Zaroubi et~al.,}{Zaroubi
  et~al.}{2012}]{Zaroubi2012}
Zaroubi S.  et~al., 2012, MNRAS,
  425, 2964

\bibitem[\protect\citeauthoryear{Zel'dovich}{Zel'dovich}{1970}]{Zel'dovichYa.B%
.1970}
Zel'dovich Y.~B.,  1970, A\&A, 5, 84

\end{thebibliography}


\appendix

\section{Error analysis}
When considering the errors on the moments of the brightness
temperature PDF we take the values measured from the simulation to be
the true statistic and then ask what happens when Gaussian random noise, 
independent but identically distributed, is introduced. The true moments
in which we are interested in measuring are given by,

\begin{equation}
  \begin{split}
    S_2&=\frac{1}{N_{\rm pix}}\sum_{i=0}^{N_{\rm pix}}(\delta T_i-\overline{\delta T})^2\,,\\
    S_3&=\frac{1}{N_{\rm pix}}\sum_{i=0}^{N_{\rm pix}}(\delta T_i-\overline{\delta T})^3\,,\\
    K_4&=\frac{1}{N_{\rm pix}}\sum_{i=0}^{N_{\rm pix}}(\delta T_i-\overline{\delta T})^4\,,
  \end{split}
\end{equation}

\noindent describing the variance, skew and kurtosis
respectively.
The brightness temperature in the output boxes is the `true' signal $\d T$ 
and its mean is $\delta \overline{T} = N_{\rm pix}^{-1}\sum_{i=0}^{N_{\rm pix}} \delta T_i$.
We assume that what is actually measured, $x$, is the linear combination of 
true signal and noise $n$, i.e. in each pixel we have 
$x_i = \delta T_i + n_i$. In this construction we imagine our
simulated maps represent some true observable signal 
in a region of the universe, which is then corrupted by 
random instrumental noise.
In doing so we only consider the effects of noise 
and are neglecting the bias induced by
sampling a restricted region of the universe.

We first considered whether there would be 
noise induced bias by testing a naive estimator
of the $m^{\rm th}$ moment given by 
$N_{\rm pix}^{-1}\sum_i^{N_{\rm pix}} (x_i - \overline{x}_i)^m$ where 
$ \overline{x}_i = N_{\rm pix}^{-1}\sum_i^{N_{\rm pix}} x_i $. We assume 
throughout the standard properties of Gaussian random noise with
a standard deviation of $\sigma$ and a mean of 0.
The even nature of Gaussian PDF returns 0 value
even moments; odd moments are given by
\begin{equation}
\langle n^m \rangle = (1) (3) (5)....(m-1)\sigma^m\,,
\end{equation}
where the angle brackets denote an estimator over noise 
realisations of their contents. This is the interpretation 
throughout as we are assuming the signal to be `true' and 
unchanging.
Some useful identities are summarised
in Table \ref{tbl:Nidents}, all exploit the assumption that
noise on different pixels is independent.
\begin{table}
    \begin{center}
      \begin{tabular}{|l|l|l|}
	\hline
        $\langle n_i\rangle$ & $=0$ & \\
        \hline

	$\langle n_in_j\rangle $ & 
        \parbox{3.3cm}{
$\langle n_i^2\rangle=\sigma_i^2\>\>\>\>\>\>\>\>\>\>(i=j)$ 
\\$\langle n_i\rangle\langle n_j\rangle=0\>\>\>\>(i\ne j)$}
        & $\bigg\}\delta_{ij}\sigma_j^2$\\  

	\hline
        $\langle n_in_j^2\rangle$ & 
         \parbox{3.3cm}{
$\langle n_i^3\rangle=0\>\>\>\>\>\>\>\>\>\>\>\>(i=j)$ \\$\langle n_i\rangle\langle n_j^2\rangle=0\>\>\> (i\ne j)$}&
        $\bigg\}0$ \\
        
        \hline
        $\langle n_i^2n_j^2\rangle$ & 
        \parbox{3.3cm}{
$\langle n_i^4\rangle=3\sigma_i^4\>\>\>\>\>\>\>\>\>\>\>\>\>\>\>\>(i=j)$
\\ $\langle n_i^2\rangle\langle n_j^2\rangle=\sigma_i^2\sigma_j^2\>\>\>(i\ne j)$}& 
        $\bigg\} (1+2\delta_{ij})\sigma_i^2\sigma_j^2$ \\

        \hline
        $\langle n_i^2n_j^3\rangle$ & 
        \parbox{3.3cm}{
$\langle n_i^5\rangle=0\>\>\>\>\>\>\>\>\>\>\>\>\>(i=j)$ \\
$\langle n_i^2\rangle\langle n_j^3\rangle=0\>\>\>(i\ne j)$} & 
        $\bigg\}0$ \\

        \hline
        $\langle n_in_j^3\rangle$ & 
        \parbox{3.3cm}{
$\langle n_i^4\rangle=3\sigma_i^4\>\>\>\>\>\>\>(i=j)$ \\
$\langle n_i\rangle\langle n_j^3\rangle=0\>\>\>(i\ne j)$} & 
        $\bigg\}3\delta_{ij}\sigma_i^4$ \\

        \hline
        $\langle n_i^3n_j^3\rangle$ & 
        \parbox{3.3cm}{
$\langle n_i^6\rangle=15\sigma_i^6\>\>\>\>\>(i=j)$\\
$\langle n_i^3\rangle\langle n_j^3\rangle=0\>\>\>(i\ne j)$} & 
        $\bigg\}15\delta_{ij}\sigma_i^6$\\
        
	\hline

      \end{tabular}
    \end{center}
      \caption{Some useful identities of the Gaussian noise assumed in our modelling of instrumental noise.}
      \label{tbl:Nidents}
\end{table}
We first consider the question of bias in the estimator, finding a bias 
is induced by our naive estimator for the variance of the 
noisy data but not for the skewness.
\begin{equation}
\begin{split}
\hat{S_2}^{\rm test} &= \frac{1}{N_{\rm pix}}\sum_{i=0}^{N_{\rm pix}} (x_i-\overline{x})^2\\
&= \frac{1}{N_{\rm pix}}\sum_{i=0}^{N_{\rm pix}} [(\delta T_i-\delta\overline{T}) + n_i ]^2\,;\\
\langle \hat{S_2}^{\rm test} \rangle &= \frac{1}{N_{\rm pix}}\sum_{i=0}^{N_{\rm pix}} (\delta T_i-\delta\overline{T})^2 + 2(\delta T_i-\delta \overline{T})\langle n_i\rangle + \langle n_i^2\rangle\, \\
&= \frac{1}{N_{\rm pix}}\sum_{i=0}^{N_{\rm pix}}(\delta T_i-\delta\overline{T})^2 + \sigma_i^2\,.
\end{split}
\end{equation}

\noindent If we assume that $\sigma_i^2 = \sigma_{\rm noise}^2$ 
for all $i$ then we can construct the following unbiased 
estimators,
\begin{equation}
\begin{split}
\hat{S_2} &=  \frac{1}{N_{\rm pix}}\sum_{i=0}^{N_{\rm pix}}(x_i-\overline{x})^2 - \sigma_{\rm noise}^2\,, \\
\hat{S_3} &=  \frac{1}{N_{\rm pix}}\sum_{i=0}^{N_{\rm pix}}(x_i-\overline{x})^3\,.
\end{split}
\end{equation}

We then get an expression for the variance of our estimators
for the $m^{\rm th}$ moment by applying 
$V_{\hat{S}_m} = \langle \hat{S}_m\hat{S}_m^{\rm T}\rangle - \langle \hat{S}_m\rangle ^ 2$. 
As we previously established unbiased estimators the second 
term on the RHS of this equation is the statistic of 
interest squared. We choose to describe our estimators 
as $\hat{S}_2$ and $\hat{S}_3$ for the variance and 
the skewness of our brightness-temperature PDFs. 
We also define $\mu_i = \delta T_i - \delta \overline{T}$ for 
clarity, where $\delta \overline{T}=N_{\rm pix}^{-1}\sum_{i=0}^{N_{\rm pix}}\delta T_i$.  The subject of our calculations $x_i-\overline{x}$ 
becomes $\mu_i + n_i$ so that $(x_i-\overline{x})^2
=(\delta T_i +n_i - \delta \overline{T})^2 
= \mu_i^2+2\mu_in_i+n_i^2$  We derive the variance of our $\hat{S}_2$ estimator as follows,
\begin{equation}
\begin{split}
V_{\hat{S}_2}=&\Bigg\langle\frac{1}{N_{\rm pix}^2}\sum_{i=0}^{N_{\rm pix}}\sum_{j=0}^{N_{\rm pix}}
[(\mu_i + n_i)^2-\sigma_{\rm noise}^2]\\
&[(\mu_j + n_j)^2-\sigma_{\rm noise}^2]  \Bigg\rangle -(S_2)^2\,.\\
\end{split}
\end{equation}

\noindent We can further multiply this expression out and move the noise 
averaging brackets inside the summation,

\begin{equation}
\begin{split}
V_{\hat{S}_2}=&\frac{1}{N_{\rm pix}^2}\sum_{i=0}^{N_{\rm pix}}\sum_{j=0}^{N_{\rm pix}}
\bigg[ \mu_i^2\mu_j^2 + 2\mu_i^2\mu_j\langle n_j\rangle + \mu_i^2\langle n_j^2\rangle\\
&+2\mu_i\langle n_i\rangle\mu_j^2 + 4\mu_i\mu_j\langle n_in_j\rangle+2\mu_i\langle n_in_j^2\rangle\\
&+\langle n_i^2\rangle\mu_j^2 + 2\langle n_i^2n_j\rangle\mu_j + \langle n_i^2n_j^2\rangle \\
&-\sigma_{\rm noise}^2
(\mu_i^2+2\mu_i\langle n_i\rangle+\langle n_i^2\rangle + \mu_j^2+2\mu_j\langle n_j\rangle + \langle n_j^2\rangle)\\
&+\sigma_{\rm noise}^4 \bigg] - (S_2)^2\,.
\end{split}
\end{equation}

\noindent Making use of the identities in Table \ref{tbl:Nidents} this reduces to 

\begin{equation}
\begin{split}
V_{\hat{S}_2}=&\frac{1}{N_{\rm pix}^2}\sum_{i=0}^{N_{\rm pix}}\sum_{j=0}^{N_{\rm pix}}
\bigg[ 
\mu_i^2\mu_j^2 
+ \mu_i^2\sigma_j^2
+ 4\mu_i\mu_j\delta_{ij}\sigma_j^2 \\
&+\sigma_i^2\mu_j^2 
+ 3\delta_{ij}\sigma_i^4+(1-\delta_{ij})\sigma_i^2\sigma_j^2\\
&-\sigma_{\rm noise}^2
(\mu_i^2
+\sigma_i^2 
+ \mu_j^2
+ \sigma_j^2)\bigg]\\
&+\sigma_{\rm noise}^4 - (S_2)^2\,.
\end{split}
\end{equation}

\noindent We can reduce this further by 
making some summation operations, 
where delta functions effectively convert all indices from 
i to j or vice versa and we get,
\begin{equation}
\begin{split}
V_{\hat{S}_2}=&
\bigg( 
S_2^2 
+ S_2 \frac{1}{N_{\rm pix}}\sum_{j=0}^{N_{\rm pix}}\sigma_j^2
+ \frac{4}{N_{\rm pix}^2}\sum_{i=0}^{N_{\rm pix}}\mu_i^2\sigma_i^2 \\
&+\frac{1}{N_{\rm pix}}\sum_{i=0}^{N_{\rm pix}}\sigma_i^2 S_2 
+ \frac{3}{N_{\rm pix}^2}\sum_{i=0}^{N_{\rm pix}}\sigma_i^4\\
&+ \frac{1}{N_{\rm pix}^2}\sum_{i=0}^{N_{\rm pix}}\sum_{j=0}^{N_{\rm pix}}\sigma_i^2\sigma_j^2 - 
\frac{1}{N_{\rm pix}^2}\sum_{i=0}^{N_{\rm pix}}\sigma_i^4\\
&-S_2 \sigma_{\rm noise}^2
-\frac{\sigma_{\rm noise}^2}{N_{\rm pix}}\sum_{i=0}^{N_{\rm pix}}\sigma_i^2 
-S_2 \sigma_{\rm noise}^2
-\frac{\sigma_{\rm noise}^2}{N_{\rm pix}}\sum_{j=0}^{N_{\rm pix}}\sigma_j^2\bigg)\\
&+\sigma_{\rm noise}^4 - (S_2)^2\,.
\end{split}
\end{equation}

\noindent Again assuming that $\sigma_i^2 = \sigma_{\rm noise}^2$ for all $i$, a multitude
of cancellation leave

\begin{equation}
V_{\hat{S}_2}= 
\frac{2}{N_{\rm pix}} (2S_2\sigma_{\rm noise}^2 
+ \sigma_{\rm noise}^4)
\,.\\
\end{equation}

An identical procedure can be used to obtain
an expression for the variance on $\hat{S}_3$:
\begin{equation}
\begin{split}
V_{\hat{S}_3}&= 
\frac{3}{N_{\rm pix}}(3\sigma_{\rm noise}^2K_4
+12S_2\sigma_{\rm noise}^4
+5\sigma_{\rm noise}^6) \,.\\
\end{split}
\end{equation}

Because we consider the normalised skew in this paper 
it is necessary to consider how the errors we have so 
far calculated propagate on to the normalised quantity. 
For the error of a function $f(X,Y)$ of two non-independent 
variables $X$ and $Y$, Taylor expansion about the expectation 
values for $X$ and $Y$ provides an approximation to the variance
on $f(X,Y)$ as a function of the errors on $X$ and $Y$,
\begin{equation}
\begin{split}
V[f(X,Y)] &\approx \left(\frac{\partial f}{\partial X}\right)^2 V[X] +  \left(\frac{\partial f}{\partial Y}\right)^2 V[Y]\\
&+2\left(\frac{\partial f}{\partial X}\right)\left(\frac{\partial f}{\partial Y} C[X,Y]\right)\,,\\ 
\end{split}
\end{equation}

\noindent where we use $V_{X}$ to denotes the variance of a quantity, here $X$, 
and $C[X,Y]=\langle XY \rangle-\langle X \rangle\langle Y \rangle$ 
to describe the covariance of two quantities, e.g. $X$ and $Y$. 
Partials are carried out fixing $X=\langle X \rangle$ 
or $Y=\langle Y\rangle$ as appropriate.
The covariance between our 
skew $\hat{S}_3$ and
variance $\hat{S}_2$ estimators can be calculated in much the 
same way as for the 
variance of each, resulting in

\begin{equation}
\begin{split}
C_{\hat{S}_2\hat{S}_3}
=& \frac{6}{N_{\rm pix}} 
S_3 \sigma^2_{\rm noise} 
\,.\\
\end{split}
\end{equation}

The variance of our estimator for the
normalised skew, $\gamma_3=\hat{S}_3/\hat{S}_2$ is
found to be
\begin{equation}
\begin{split}
V_{\gamma_3} \approx \frac{1}{(S_2)^2}V_{\hat{S}_3}
+\frac{(S_3)^2}{(S_2)^4}V_{\hat{S}_2}
-2\frac{S_3}{(S_2)^3}C_{\hat{S}_2\hat{S}_3}\,,\\ 
\end{split}
\end{equation}

\noindent and for $\gamma'_3=\hat{S}_3/(\hat{S}_2)^{3/2}$ we have
\begin{equation}
\begin{split}
V_{\gamma'_3} \approx \frac{1}{(S_2)^3}V_{\hat{S}_3}
+\frac{9}{4}\frac{(S_3)^2}{(S_2)^5}V_{\hat{S}_2}
-3\frac{S_3}{S_2^{4}}C_{\hat{S}_2\hat{S}_3}\,.\\ 
\end{split}
\end{equation}

\bsp

\end{document}